\documentclass[10pt,aps,prx,twocolumn,superscriptaddress,showpacs]{revtex4-1}
\usepackage{amsmath}
\usepackage{amsfonts}
\usepackage{graphicx} 
\usepackage{float}
\usepackage{dcolumn}
\usepackage{bm}
\usepackage{color}

\begin{document}

\newcommand{\Itilde}{\widetilde{I}}
\newcommand{\ptilde}{\widetilde{p}}
\newcommand{\li}[1]{ {\color{blue} #1}}
\newcommand{\be} {\begin{equation}}
\newcommand{\ee} {\end{equation}}

\title{Full counting statistics of time of flight images}

\author{Izabella Lovas}
\affiliation{MTA-BME Exotic Quantum Phases ``Momentum'' Research Group and
Department of Theoretical Physics, Budapest University of Technology and Economics, 1111 Budapest, Hungary}
\author{Bal\'azs D\'ora}
\affiliation{MTA-BME Exotic Quantum Phases ``Momentum'' Research Group and
Department of Theoretical Physics, Budapest University of Technology and Economics, 1111 Budapest, Hungary}
\author{Eugene Demler}
\affiliation{Physics Department, Harvard University, Cambridge, Massachusetts 02138, USA}
\author{Gergely Zar\'and}
\affiliation{MTA-BME Exotic Quantum Phases ``Momentum'' Research Group and
Department of Theoretical Physics, Budapest University of Technology and Economics, 1111 Budapest, Hungary}

\begin{abstract}

Inspired by recent advances in cold atomic systems and non-equilibrium physics,
we introduce a novel characterization scheme, the time of flight full counting statistics. 
   We benchmark this method on an interacting one dimensional Bose gas, and show that 
there the time of flight image  displays 
several universal regimes.
Finite momentum fluctuations are observed at larger distances, where   a crossover  from exponential to Gamma distribution occurs
upon decreasing  momentum resolution. Zero momentum particles, on the other hand, obey a Gumbel distribution in the weakly interacting limit,  characterizing the quantum
fluctuations  of the former quasi-condensate. 
Time of flight full counting statistics is demonstrated  to capture (pre-)thermalization 
processes after a quantum quench, and can be useful for characterizing   exotic quantum states  
such as many-body localized systems or models of holography.
\end{abstract}



\maketitle

\section{Introduction}

 One of the fundamental principles of modern theory of strongly correlated many-body systems is emergent universal behavior. For example, in the vicinity of a thermal phase transition, one finds universal behavior of correlation functions determined by the
nature of the transition but not the microscopic details \cite{pt,cardy,kadanoff,sachdev}. Close to criticality, the behavior of correlation functions is just 
determined by the dimensionless ratio of the system size and one emergent 
lengthscale: the correlation length \cite{cardy,sachdev}. This statement is expected to hold beyond two point
correlation functions. Higher order correlation functions and distribution functions should also obey hyper-scaling property: they are universal functions of the the system size to the correlation length. While hyperscaling has been well studied theoretically \cite{cardy,sachdev},
it has not been observed in experiments so far. 

In quantum systems we expect manifestations of emergent universality to be even stronger. For example, we expect that a broad class of one dimensional quantum systems can be described by a universal Luttinger
theory \cite{Giamarchi,Cazalilla,tsvelik,haldane,delft}. This powerful approach demonstrates that long distance correlation functions as well as low energy collective modes are described by a universal theory which is not sensitive to details of underlying microscopic Hamiltonians. This powerful paradigm of
universality has been commonly discussed in the context of two point correlation functions, such as probed by scattering and tunneling experiments \cite{HBTBEC,conductivity,tarucha,goni,meirav,tennant}.

 In principle, to fully characterize these in or out of equilibrium quantum states at every instant, 
one should reconstruct them by performing   Quantum State Tomography. 
In practice, however, quantum state tomography is restricted  to tiny quantum systems \cite{tomo}.
The most complete information on the many-body wave function can be  obtained 
  through investigating the \emph{full distribution} of some properly chosen physical 
  observables~\cite{hofferberth,levitov,nazarov,ensslin,lu,pekola,silva,batalhao}.
 Observing  universality in these distribution functions  would therefore be a 
 direct and striking demonstration of the universal nature of the entire many-body state and emergent universality.

Unfortunately, in traditional solid state systems, experimental studies of such distribution functions are extremely challenging. 
Most of experimental techniques rely either on averaging over many 1d systems, such as in a crystal containing many 1d
systems \cite{tennant,magishi,bourbonnais}, or on long time averaging such as in STM experiments \cite{stm1,stm2,stm3}. 
As a result, 
no theoretical work has been done on understanding universality classes of distribution functions of observables in quantum systems. 

Recent progress with ultracold atoms, however, makes
it possible to perform experiments that look like textbook classical measurements of quantum mechanical wavefunctions on individual 1d systems \cite{ZwergerReview}. By collecting a histogram of single shot results one can obtain full distribution functions. In particular, quasi-one dimensional gases have provided an interesting test-ground  to realize
and test low dimensional quantum field theories~\cite{lowD}.
In a peculiar setup, a pioneering series of sophisticated experiments was performed~\cite{hofferberth,gring,schmiedmayer2} to access the  probability distribution function (PDF)~\cite{Polkovnikov,Imambekov2,Zwerger} of matter-wave interference fringes of a coherently split one-dimensional Bose gas  and to gain deeper insight into  phase correlations.

Here we propose that even the most wide-spread and extensively used \emph{standard} Time of Flight (ToF) images contain a lot more 
precious  information -- never exploited so far, which can be  extracted  and used to characterize the quantum state observed. 
In particular, we propose to study  the full distribution function of Time of Flight images, 
a procedure we dubbed   \emph{Time of Flight Full Counting Statistics}  to parallel the  method used
in nanophysics  \cite{levitov,nazarov,ensslin,lu,pekola}.
ToF  imaging is in fact probably \emph{the} most wide-spread tool to investigate cold atomic systems \cite{ZwergerReview,ToF,Bouchoule,Mott}, and a wide range of other, more sophisticated experimental techniques like  Bragg spectroscopy 
or matter-wave interference are also based on ToF measurements. 
In a ToF experiment with quasi one and two dimensional systems, atoms quickly cease to interact  after being released from a trap, and therefore  their position after some time is directly 
 proportional to their momenta in the initial quantum state. 
 ToF images thus  picture the momentum distribution of the atoms 
 in the initial interacting state (see Fig.~1).
They  contain, however, a lot more information than just  the average intensities or their correlations \cite{footnote6} they contain the \emph{full} probability distribution function (PDF) of particles at each momentum, which is expected to reflect the universal behavior of low dimensional quantum systems 
or critical states. In this work, we concentrate on this so far unexploited information, accessible in 
a wide range of experimental settings for many experimental groups.

\begin{figure}[b!]
\includegraphics[width=\columnwidth]{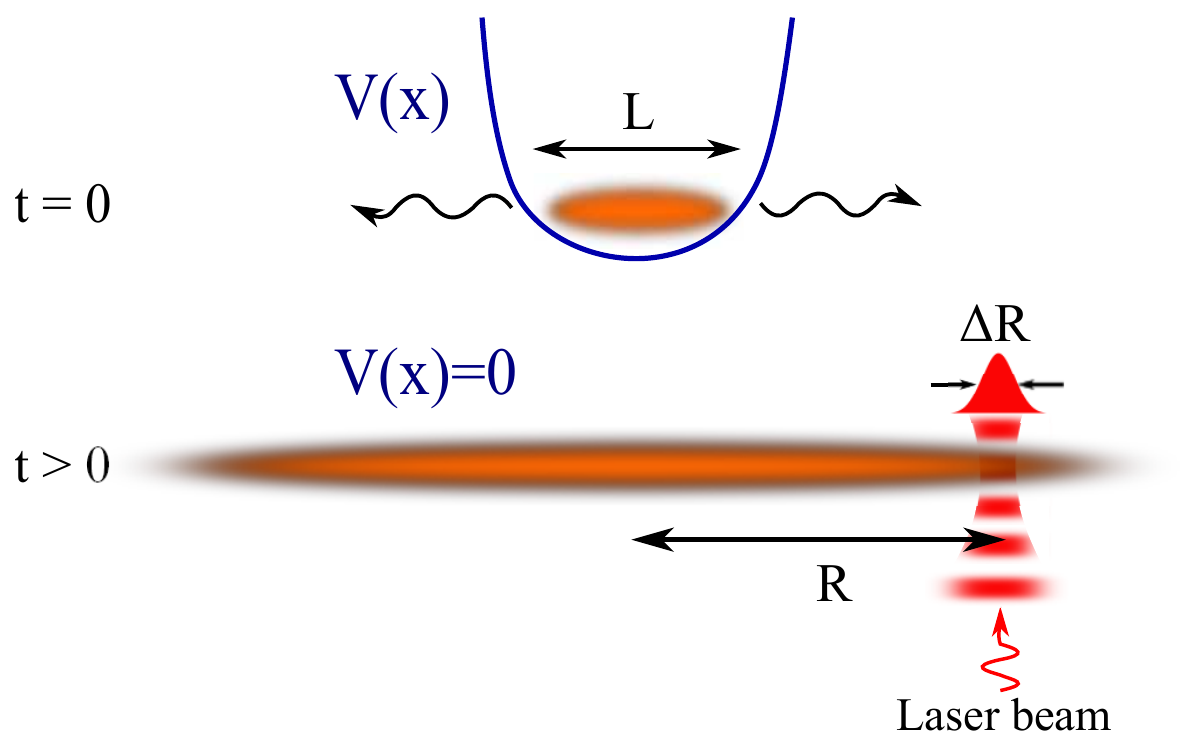}
\caption{Sketch of ToF experiment with quasi-one-dimensional Bose gas. At $t=0$ the atoms, initially confined to a tube of length $L$, are
released form the trap. Interactions between the particles are typically short ranged, and become quickly negligible due to the rapid expansion in transverse direction (not shown here). After propagation time $t$, the density profile of the expanded cloud is investigated by taking an absorption image
at position $R$ with a laser beam of waist $\Delta R$. Atoms expand freely after  release from the trap, and 
the distribution of the measured intensity provides direct information on the  structure of the initial quantum state.}\label{fig:sketch}
\end{figure}

 To demonstrate this approach,  we analyze  the fingerprints of abundant quantum fluctuations on  one dimensional interacting quasi-condensates,
 and  determine the  \emph{complete} distribution of the time of flight image. The particular  setup considered is sketched in Fig. \ref{fig:sketch}: a one dimensional Bose gas, confined to a tube of length $L$,
is suddenly released from the trap. Due to the rapid expansion in the tightly confined directions (not shown in  Fig. \ref{fig:sketch}), the interactions become quickly negligible, and it is enough to consider a free, one dimensional propagation along the longitudinal axis \cite{focusing1}.  After expansion time $t$, the density profile is imaged by a laser beam at position $R$, which
measures the integrated density of particles within the  spotsize  of the laser, $\Delta R$,
\begin{equation}\label{def}
\hat{I}_{R,\Delta R}(t)\equiv\int_{-\infty}^{\infty} {\rm d}x\, e^{-(x-R)^2/(2\Delta R^2)}\hat{\psi}^\dagger(x,t)\hat{\psi}(x,t).
\end{equation}
Here $\hat{\psi}(x,t)$ denotes the bosonic field operator, and we assumed~\cite{laser}
 a Gaussian laser intensity profile~\cite{footnote1}.
Since bosons propagate freely during the ToF expansion,
Eq. \eqref{def} provides information on the correlations in the initial state of the system at time $t=0$.
In particular, for  $R\gg L$ and $\Delta R\ll R$, the measured intensity  can be  interpreted as
the number of particles $\hat N_p$ with a given initial momentum, $p=m R/t$ \cite{ToF}. Let us note that the same information about momentum correlations can also be obtained by performing another experimental procedure, the focusing technique ~\cite{focusing0,focusing1,focusing} (see  Appendix~\ref{app:focusing}). As we discuss later, apart from minor corrections, our results apply for this type of measurement as well \cite{footnote4}, which offers, however, a more accurate approach to measuring distributions in momentum space than the usual ToF technique.

 We  determine  the full distribution  of the  operator $\hat{I}_{R,\Delta R}$, and show
that it  contains important  information on the quantum fluctuations of the condensate, leading to the emergence of several universal distribution functions.
Analysing first the image of a $T\approx 0$ temperature condensate, we show that
 intensity distributions at finite momenta follow Gamma distribution, and reflect squeezing. The signal of zero momentum particles is,
on the other hand,  shown to follow a Gumbel distribution in the weakly interacting limit, a characteristic universal distribution of extreme value statistics, and reflecting
large correlated particle number fluctuations of the quasi-condensate. We also extend our  calculations to finite temperatures and show 
that the predicted  Gumbel distribution should  be observable at realistic  temperatures for typical system parameters.
Then we study the image of the condensate after a quench, and show
how thermalization of the condensate manifests itself as a crossover to an - also universal - exponential distribution in the time of flight  full counting statistics.

\section{Theoretical framework}
 

To reach our  main goal and  to determine the full distribution of  $\hat{I}_{R,\Delta R}(t)$ for a one dimensional 
interacting Bose gas, we shall make use of  Luttinger-liquid theory, \cite{Imambekov1,Imambekov2}
and compute all moments  of $\hat{I}_{R,\Delta R}$  to show that for  long  times of flight
and large enough distances 
\begin{equation}\label{pdfdef}
\langle  \hat{I}^n_{R,\Delta R}\rangle (t)\to \int_0^\infty{\rm d}I \; I^n\,W_{p}(I)\;
\end{equation}
with $n$ positive integer.
The function   $W_p(I)$  can be viewed  as the probability distribution function (PDF) of 
the intensity, measuring the number of particles $N_p\sim I$ with momentum $p= m R/t$.
Notice that the function $W_{p}(I)$ depends  implicitly on the time of flight as well as on 
the momentum resolution $\Delta p$, suppressed for clarity in Eq.~\eqref{pdfdef}.
 %

Luttinger-liquid theory describes
the low energy properties of quasi-one-dimensional bosons~\cite{Cazalilla}  
as well as a wide range of one-dimensional systems~\cite{Giamarchi}. 
Long wavelength excitations of a Luttinger liquid are collective bosonic modes, 
described in terms of  a phase field, $\hat{\phi}(x)$ \cite{Giamarchi}.
For  quasi-condensates,   the field operator $\hat{\psi}(x)$  is directly related to this phase operator~ \cite{Giamarchi}
\begin{equation}
\label{densphase}
\hat{\psi}(x)\approx \sqrt{\rho}\;e^{i\hat{\phi}(x)},
\end{equation} 
with  $\rho$  the average density of the quasi-condensate. Fluctuations of the density generate dynamical phase fluctuations, 
 described by a simple Gaussian action \cite{Giamarchi,ZwergerReview},
\begin{equation}\label{action}
S=\dfrac{K}{2\pi}\int {\rm d}t\int {\rm d}x\left(\dfrac{1}{c}\left(\partial_t\phi\right)^2-c\left(\nabla\phi\right)^2\right),
\end{equation}
that involves the sound velocity of bosonic excitations, $c$, and the Luttinger parameter, $K$. 
The dimensionless parameter $K$ characterizes the strength of the interactions: 
for hard-core bosons $K\to1$, corresponding to the so-called Tonks-Girardeau limit~\cite{Girardeau, TGexperiment}, 
while for weaker repulsive interactions $K>1$, with $K\rightarrow\infty$ corresponding to the non-interacting limit~ \cite{Giamarchi}. 
The connection between the parameters $c$ and $K$ and the microscopic  parameters is model dependent. 
For a weak repulsive Dirac-delta interaction, $V(x-x^\prime)=g\,\delta(x-x^\prime)$, both are determined  by 
 perturbative expressions \cite{Cazalilla}
\begin{equation}\label{perturb}
c\cong\sqrt{\dfrac{g\rho}{m}},\quad K\cong\dfrac{\hbar\pi\rho}{m c}=\hbar\pi\sqrt{\dfrac{\rho}{m g}},
\end{equation}
with $\hbar$  the Planck constant. 

To  evaluate the moments of the operator $\hat{I}_{R,\Delta R}(t)$, we  first observe that the interactions between the atoms 
become quickly negligible once  the confining potential is turned off and the atoms start to expand. 
 Therefore the fields $\hat \psi(x,t)$ evolve almost freely in time for times $t>0$, with a  time evolution described  
by  the  Feynman  propagator, $G(x,t)\sim e^{i m x^2/(2\hbar t)}/\sqrt{i \,t}$, 
\begin{equation}\label{eq:propagator}
\hat\psi(x,t) = \int {\rm d}x'\, G(x-x',t)\hat \psi(x')\, .
\end{equation}
For large times, and points far away from the initial position of the condensate one finds that 
$\hat\psi(x,t)$ is approximately equal to the Fourier transform of the field  $\hat \psi_p$ at a momentum $p= m x/t$.
This relation becomes exact if, instead of  a simple time of flight experiment, one  uses the previously mentioned focusing 
technique (see  Appendix~\ref{app:focusing}), allowing to reach much better resolutions ~\cite{focusing0,focusing1,focusing}.

Applying the  representation Eq.~\eqref{densphase} and the   Gaussian action Eq.~\eqref{action}, we can  evaluate 
 $\langle  \hat{I}_{R,\Delta R}^n(t)\rangle$ in  any moment~\cite{Imambekov1,Imambekov2}, 
 and construct the intensity distribution $W_p(I)$.
Using open boundary conditions for the phase operator we obtain, e.g. 
\begin{equation}\label{w}
W_p(I)=\int\int_{-\infty}^{\infty}\prod_{j}\dfrac{{\rm d}\tau_{j}\,e^{-\tau_{j}^2/2}}{\sqrt{2\pi}}\delta\left(I-\dfrac{N\Delta \widetilde{p}}{\sqrt{2\pi}}g\left(\lbrace \tau_{j}\rbrace\right)\right),
\end{equation}
with ${j}=1,2,\dots$ labeling the auxiliary variables $\tau_j$ and the function $g\left(\lbrace \tau_{j}\rbrace\right)$ determined by the double integral
\begin{align}\label{gtf}
&g\left(\lbrace \tau_j\rbrace\right)=\int\int_{-1/2}^{1/2}{\rm d}u\,{\rm d}v\, e^{ - \Delta \widetilde{p}^2(u-v)^2/2+i\,\widetilde{p}(u-v)\left(1-\frac{u+v}{2 R/L}\right)}\cdot\nonumber\\
&\qquad\exp\left(i\sum_{j}\tau_{j}\;\frac{e^{-\xi_h\pi {j}/(2L)}}{\sqrt{K\,{j}}}\left\lbrace\cos\left(\pi \,{j} \,u+\frac{{j}\,\pi}{2}\right)\right.\right. \nonumber\\
&\qquad\qquad\qquad\quad-\left.\left.\cos\left(\pi\, {j}\, v+\frac{{j}\,\pi}{2}\right)\right\rbrace\right).
\end{align}
The derivation of Eqs.~\eqref{w} and \eqref{gtf} is detailed in Appendix \ref{app:pdf}.
The healing length  $\xi_h\equiv \hbar/(m c)$ here serves  as a short  distance cutoff~\cite{footnote2}, 
 $N=L\rho$ denotes the total number of particles, and we introduced the  dimensionless time of flight momentum  and its resolution
\begin{equation}\label{pdp}
\widetilde{p}\equiv\dfrac{mR}{t}\dfrac{L}{\hbar}\;\;\;,\quad\Delta \widetilde{p}\equiv\dfrac{m\Delta R}{t}\dfrac{L}{\hbar}\;,
\end{equation}
both measured in units of $\hbar/L$. 

We note that the intensity measured in a focusing experiment also follows a distribution of the form of Eq. \eqref{w}, apart from  a small change in the function $g\left(\lbrace \tau_j\rbrace\right)$.  As discussed in Appendix~\ref{app:focusing}, in a focusing experiment Eq.~\eqref{eq:propagator}  yields just the Fourier transform  of the field $\hat\psi$,  and the real space coordinates $R$ and $\Delta R$ are 
directly proportional to the dimensionless momenta, $\tilde{p}$ and $\Delta\tilde{p}$.
As a technical consequence,  the term $\exp(-i\tilde{p}(u^2-v^2)L/(2R))$ is absent from the integral giving $g\left(\lbrace \tau_j\rbrace\right)$, 
but for a given $\tilde{p}$ and $\Delta\tilde{p}$, the shape of distribution is hardly affected by this minor modification in the relevant limit, $R\gg L$. 
Therefore all the results presented below apply also for  intensities measured by the refocusing method.

\section{Equilibrium quantum fluctuations}

We  evaluated  Eqs.~\eqref{w} and \eqref{gtf} by performing   classical Monte Carlo simulations.
Already the expectation values, $\langle\hat{I}_{R}\rangle$
 carry valuable information,  since they  account for the size of  interaction induced quantum (or thermal) 
 fluctuations of  bosons  with momentum $p=mR/t$.  They are proportional to 
$\langle \hat N_p\rangle $ and   to the corresponding momentum dependent 
effective temperatures.  The momentum and temperature dependence of 
$\langle \hat N_p\rangle $ has been studied theoretically~\cite{Giamarchi}
 and experimentally~\cite{TGmomentum,1/k4}
 in detail (see also the following  subsections and Appendix~\ref{app:expectation_value}). 
In an infinitely long Luttinger liquid, in particular,  
$\langle \hat N_p\rangle$  falls of as  $\sim 1/|p|^{1-1/2K}$
at $T=0$ temperature, while at finite temperatures its value 
depends on $p$: For small momenta it saturates to a constant  proportional to $1/T^{1-1/2K}\approx 1/T$, while 
at large momenta the power law behavior is recovered. For weak interactions, the cross-over between these two regimes occurs through a regime, where a power law behavior is observed with a modified exponent (see Appendix~\ref{app:expectation_value}). 

The average being well understood, here we  concentrate on the \emph{shape} of the  full intensity 
distribution. Therefore, 
we introduce  the normalized intensity
\begin{equation*}
\widetilde{I}=\hat{I}_{R,\Delta R}/\langle\hat{I}_{R,\Delta R}\rangle,
\end{equation*}
and determine the corresponding distribution function  $\widetilde{W}_p(\widetilde{I})$. 
The intensity distributions for $p=0$  and  for typical  $p\ne 0$
exhibit drastically  different characters;  the  zero momentum intensity is just associated with particles in 
the  quasi-condensate,  while intensities corresponding  to $p\ne0$ reflect quantum fluctuations to 
states of momentum   $p$.  We shall therefore discuss these separately.

\begin{figure}[t!]
\includegraphics[width=0.9\columnwidth]{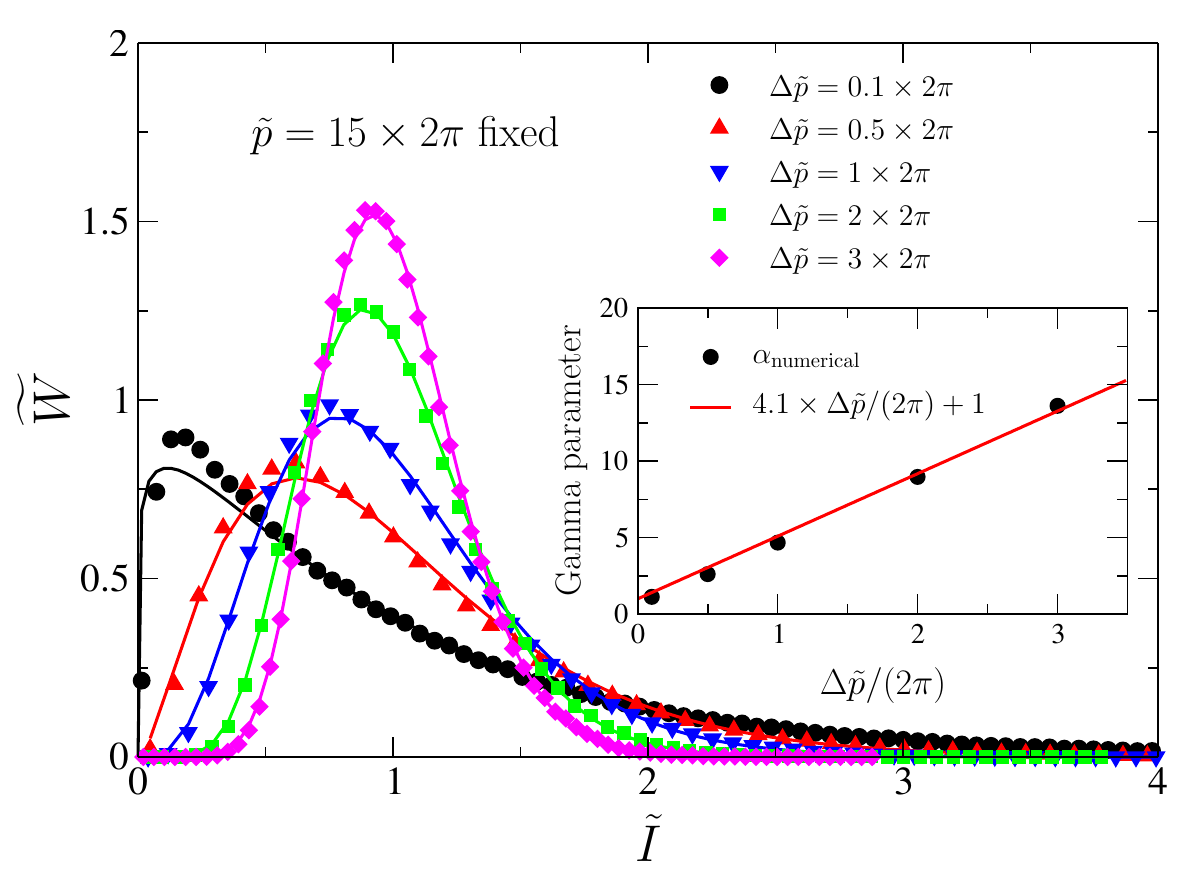}
\caption{Distribution  of normalized intensity $\widetilde{I}$ (symbols) at $T=0$ temperature, plotted for different momentum 
resolutions,  $\Delta \widetilde{p}$. We used  $K=10$, $\widetilde{p}=15\times 2\pi$ and $\xi_h/L=0.002$. Solid lines are fits with the 
Gamma distribution from Eq.~\eqref{Gamma}. The distribution smoothly evolves from exponential to Gamma as $\Delta p$ increases, and
reflects  the two-mode squeezed structure of the Bogoliubov ground state in momenta $p$ and $-p$. 
Inset: parameter of the fitted Gamma distribution $\alpha$ as a function of $\Delta \ptilde$. 
%
%
\label{fig:gamma}
}
\end{figure}

\subsection{Zero-temperature  intensity distribution  of finite momentum particles}
\label{sub:T0}

Let us first discuss  the intensity distribution  of finite  momentum particles, $p\ne0$,  
at $T=0$ temperature, allowing us to take a glimpse at  the structure of   interaction-generated quantum fluctuations.
Fig.~\ref{fig:gamma} shows the  typical  structure of the distribution function $\widetilde{W}_p(\widetilde{I})$  
for a moderate  Luttinger parameter $K=10$  for various momentum resolutions  $\Delta\widetilde{ p}$. 
The shape of   $\widetilde{W}_p(\widetilde{I})$  has a strong dependence on the 
 resolution $\Delta \ptilde $, and is well described by a Gamma distribution 
\begin{equation}\label{Gamma}
\widetilde{W}_{p\ne0}(\widetilde{I})\approx \dfrac{\alpha^\alpha}{\Gamma(\alpha)}\,\Itilde^{\alpha-1}\,e^{-\alpha\,\Itilde}.
\end{equation}
The parameter $\alpha$ here incorporates  the momentum resolution, $\Delta\ptilde$, and 
increases linearly with  it  (see inset of Fig~\ref{fig:gamma}). 
For good resolutions   $\alpha\approx 1$, 
 an exponential distribution is recovered, 
$$
\widetilde{W}_{p\ne0}(\widetilde I)\approx e^{-\widetilde{I}},\phantom{nnn} \text{for }\phantom{nn} {\Delta \widetilde p\ll  2\pi}.
$$


These observations can be understood in terms of  the Bogoliubov approximation \cite{Bogoliubov}, valid for weak interactions and short system sizes. 
For small sizes of the laser spot, i. e. $\Delta \ptilde \ll 2\pi$, the intensity, Eq.~\eqref{def} can be interpreted as the number 
of particles with dimensionless wave number $p=m R/t$.
%
%
%
The Bogoliubov ground state has a two-mode squeezed structure, 
i.e.,   particles with momenta $p$ and $- p$ are always created in pairs, implying perfect correlations at the operator level, 
 $\hat{N}_p=\hat{N}_{-p}$. 
This two-mode squeezed structure gives rise to  a geometric distribution for the particle number $\hat{N}_p$ \cite{2modesqueez}, and the exponential intensity distribution observed 
 is just the continuous version of this geometric distribution.

Moreover, Bogoliubov theory predicts vanishing correlation between nonzero momenta $|p|\ne |p'|$ \cite{Bogoliubov}.  Therefore, 
 the total number of particles in a given momentum window  $\Delta p$
  can be viewed as the sum of $\sim \Delta \ptilde/2\pi$ independent, exponentially distributed random variables, with 
  approximately equal expectation values~\cite{Bogoliubovmomentum}
\begin{equation}\label{Bogoliubov}
\langle\hat{N}_p\rangle\approx\frac{\rho\, \hbar\,\pi}{2 K |p|}\,.
\end{equation}
The Gamma distribution with a parameter $\alpha\propto \Delta\ptilde$  thus arises as the weighted sum of independent exponential variables. 
The precise prefactor here depends on the shape of the intensity profile in Eq.~\eqref{def}. For a Gaussian profile 
we find $\alpha\approx 4.1\;\Delta\ptilde/(2\pi)$, while other profiles amount in other numerical prefactors of ${\cal O}(1)$. 
%
Though the Bogoliubov approach has only a limited range of validity,
a similar  crossover from exponential to Gamma distribution persists even for strong interactions (see Appendix \ref{app:suppfig}).

\subsection{Quasicondensate  distribution at $T=0$ temperature}

\begin{figure}[t]
\includegraphics[width=0.8\columnwidth]{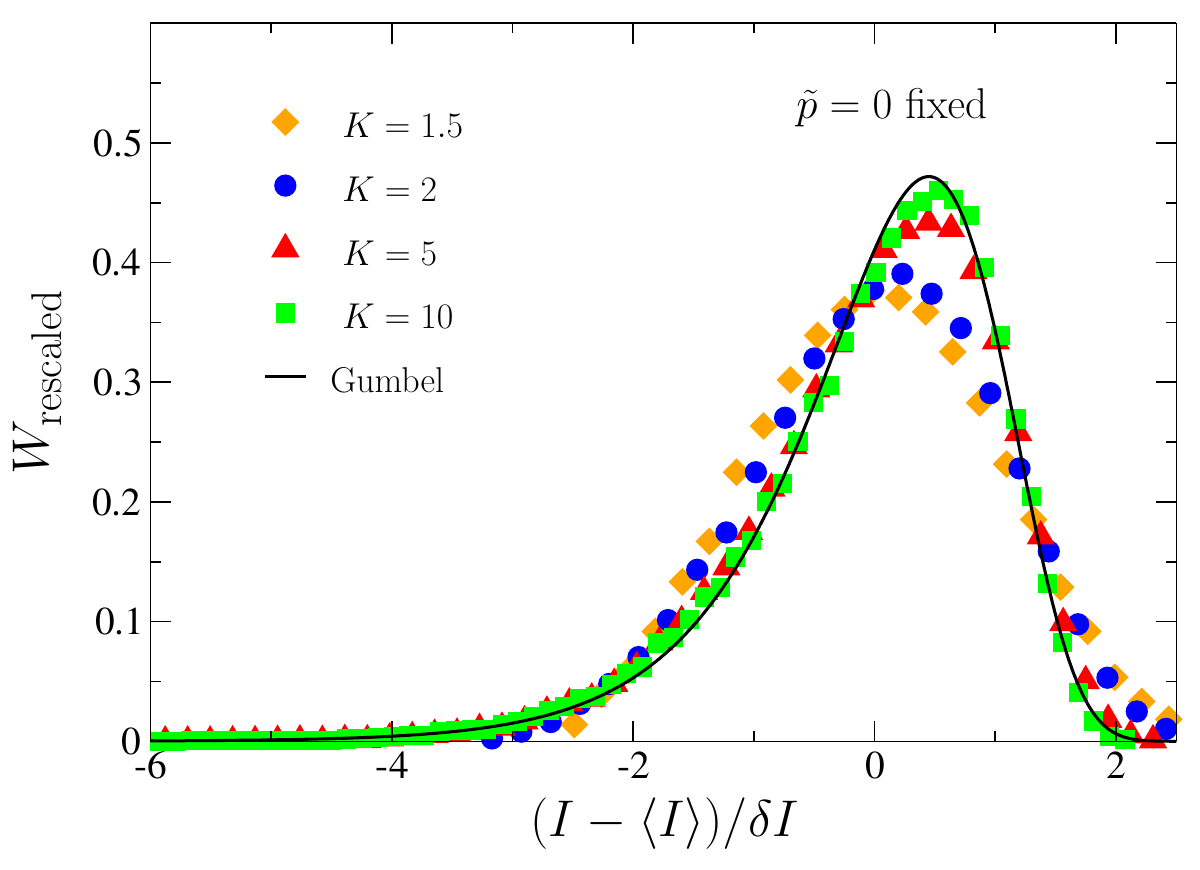}
\caption{PDF of the normalized variable $(\hat{I}-\langle\hat{I}\rangle)/\delta I$ for  $p=mR/t=0$,  plotted for different Luttinger parameters $K$, 
with $\delta I$ referring to standard deviation. We used periodic boundary conditions  to compare to analytical results. For weak interactions ($K\gg1$) 
the PDF converges to Gumbel distribution, Eq.~\eqref{gumbel} (solid line), also predicted  by a particle number preserving 
Bogoliubov approach. We used $\Delta\widetilde{p}=0.1\times 2\pi$ and $\xi_h/L=0.002$.
}\label{fig:k0}
\end{figure}

 Let us now turn to the zero-momentum distribution, corresponding to the number of particles in the quasi-condensate,  
 and exhibiting  a completely different behavior, 
 shown in  Fig. \ref{fig:k0}. 
   The distribution, plotted for different interaction strengths $K$, converges quickly  to a so-called Gumbel distribution 
   as $K$ increases. This distribution, arising frequently in extreme value statistics \cite{Gumbel}, is expressed as 
\begin{equation}
\label{gumbel}
W_{\rm Gumbel}(\Itilde)=\dfrac{\pi}{\sqrt{6}}\;{\rm exp}\left(\dfrac{\pi}{\sqrt{6}}\,\Itilde -\gamma-{\rm exp}\left\lbrace\dfrac{\pi}{\sqrt{6}}\,\Itilde-\gamma\right\rbrace\right),
\end{equation}
with $\gamma\approx 0.5772 $ the Euler constant.

We can prove that the extreme value distribution \eqref{gumbel} follows from particle number conservation combined with 
 the fact that $\hat N_{p\ne0}$ display  exponential distributions with expectation values  $\langle\hat N_{p\ne0}\rangle\sim 1/|p|$.
Particle number conservation relates the fluctuations of 
the number  of particles in the condensate, $\hat N_0$ with those of  $p\ne 0$ particles,    $\hat{N}_{0}=N-\sum_{p\neq 0}\hat{N}_p$.
 This can be achieved within the particle number preserving Bogoliubov approach of Ref.~\cite{Castin}
 by performing a second order expansion in the bosonic fluctuations.
As discussed above, all finite momentum particle numbers $\hat{N}_{p\ne0}$ exhibit 
  exponential distributions with expectation values $\sim 1/|p|$.   Therefore, as we show in 
Appendix~\ref{app:gumbel},   the distribution of the sum $\sum_{p\neq 0}\hat{N}_p$ can be rewritten analytically, and  
reexpressed as the \emph{maximum} of a large number of independent, identically distributed exponential random variables, leading to 
 the observed Gumbel distribution.   

For strong interactions $K\sim 1$, the zero-momentum distribution starts to deviate form the Gumbel distribution, Eq. \eqref{gumbel}, considerably. However, the observed distribution is still universal in the sense that it does not depend on the momentum cutoff, and remains unchanged if we consider a Bogoliubov spectrum instead of the linear dispersion relation of a Luttinger liquid.

\subsection{Joint probability distribution}
 
Similar to the full distribution function, $W_p(I)$, we can generalize   usual multipoint correlation functions and define the joint distribution 
  $W_{p_1,p_2,\dots}(I_1,I_2,\dots)$, corresponding to measuring the intensities $\bigl \{ \hat{I}_{R_1}, \hat{I}_{R_2},\dots\bigr\}$
  at positions $R_i = p_i t/m$. More formally, in analogy with Eq. \eqref{pdfdef}, the joint distribution function $W(I_{R_1},I_{R_2})$ can be defined through the moments of the variables $\hat{I}_{R_1}$ and $\hat{I}_{R_2}$,
\begin{equation}\label{jointpdf}
\langle  \hat{I}^{n_1}_{R_1}\hat{I}^{n_2}_{R_2}\rangle (t)\to \int_0^\infty{\rm d}I \; I_1^{n_1}I_2^{n_2}\,W(I_1,I_2),
\end{equation}
for any positive integers $n_1$ and $n_2$.  The previous calculations can be extended to compute these  probability distributions with little effort
(see Apendix~\ref{app:jointpdf} for details).  
Without analysing them in detail, here we just briefly  discuss the joint distribution function of the 
of $p=0$ and $p\ne0$ modes, providing further evidence
 for the role of particle number conservation behind the emergent  extreme value statistics.

\begin{figure}[t!]
\includegraphics[width=0.7\columnwidth]{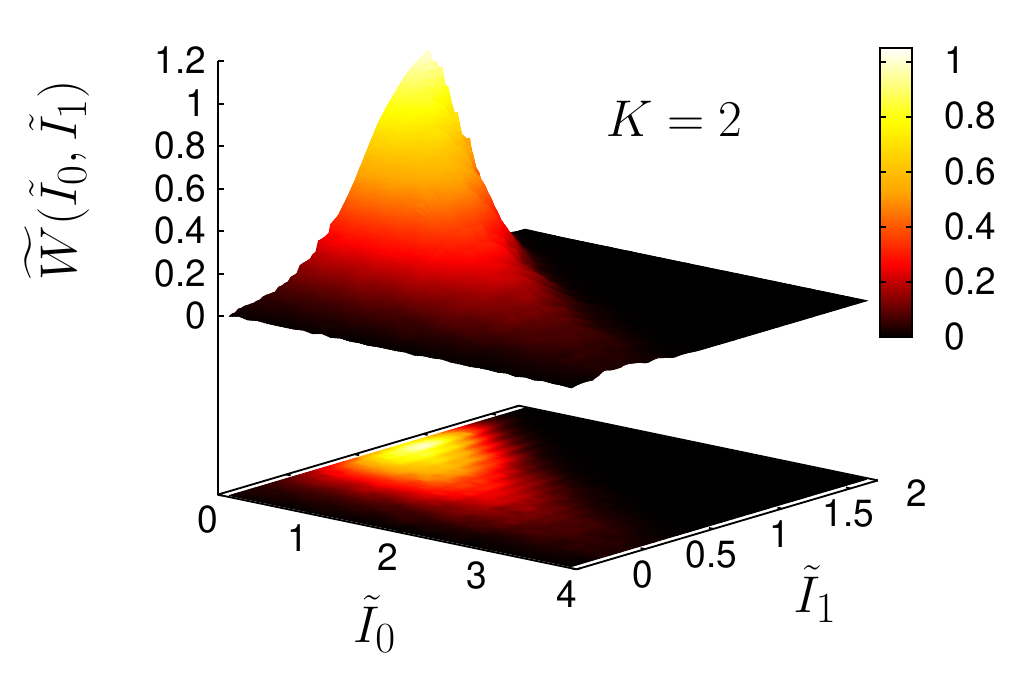} \\
\includegraphics[width=0.7\columnwidth]{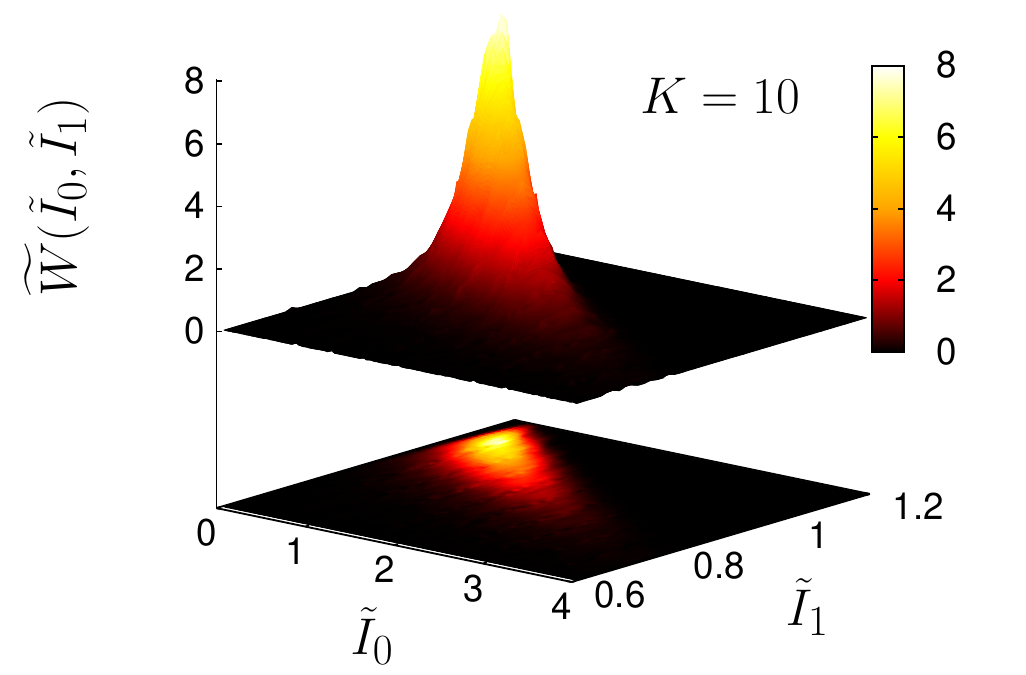} \\
\caption{Joint PDF of signals $\widetilde{I}_0$ and $\widetilde{I}_1$, evaluated for dimensionless momenta  $\ptilde_0=0$ and $\ptilde_1=2\pi$, 
for strong ($K=2$) and weak ($K=10$) interactions at $T=0$ temperature. The anticorrelation, observable for any interaction strength, reflects particle number conservation. Particles with non-zero wave numbers $p_1$, leave holes behind in the quasi-condensate. 
}
\label{fig:k0_joint_PDF}
\end{figure}

The distribution of 
 the normalized variables $\widetilde{I}_0$ and $\widetilde{I}_1$, corresponding to dimensionless momenta $\ptilde_0=0$ and $\ptilde_1=2\pi$ is 
plotted in  Fig. \ref{fig:k0_joint_PDF} for strong ($K=2$) and weak ($K=10$) interactions. 
The wave number resolution  was  chosen to be  
such that  particles contributing to 
the signals ${I}_0$ and ${I}_1$ have well defined momenta. 
The joint PDFs reveal strong anticorrelation between the intensities $I_0$ and $I_1$, interpreted as particle numbers  $N_{0}$ and $N_{1}$, for all interaction strengths, 
persisting for higher values of $p_1$. 
%
%
Anticorrelations manifest in the fact that the joint PDF is sharply peaked around the line $\widetilde{I}_0+\widetilde{I}_1={\rm const.}$, implying that a high intensity $\widetilde{I}_0$ is typically accompanied by a low signal  $\widetilde{I}_1$. 
The origin of these anticorrelations is 
particle number conservation:
a particle with non-zero wave number $p_1$, removed from the quasi-condensate, leaves a 'hole' behind, eventually appearing as
anticorrelation in the joint PDF of $I_{1}$ and $I_{0}$. 
%

\subsection{Finite temperature effects and thermal depletion of the quasi-condensate} 
\label{sub:finiteT}

So far we  focused on the limit of $T=0$ temperature. At finite temperatures, modes with energies 
$E =p \,c \lesssim k_B T$ get thermally excited and, at some point, destroy the quasi-condensate.  
As we show now, this thermal depletion of the quasi-condensate is controlled by the   dimensionless temperature 
\begin{equation}\label{Tcrossover}
\widetilde{T}=\dfrac{1}{K}\dfrac{k_B T}{\Delta} ,
\end{equation} 
with $\Delta = h\,c /L$  the 'level spacing', i.e.  the typical separation of sound modes in a condensate of 
size $L$.   

Fig.~\ref{fig:FiniteTDist} displays  the intensity distribution of the zero-mode, derived in  Appendix \ref{app:pdf},
as a function of $\widetilde{T}$ 
for  experimentally  relevant parameters~\cite{experiment,footnote5}. 
%
%
The PDF retains the characteristic shape of a Gumbel distribution for realistic but small temperatures,
$\widetilde{T}\lesssim 1$, though the distribution broadens with increasing temperature. 
 At temperatures $\widetilde{T}\gtrsim 1$, however, the PDF  
turns quickly into an exponential distribution.

\begin{figure}[b!]
\includegraphics[width=\columnwidth]{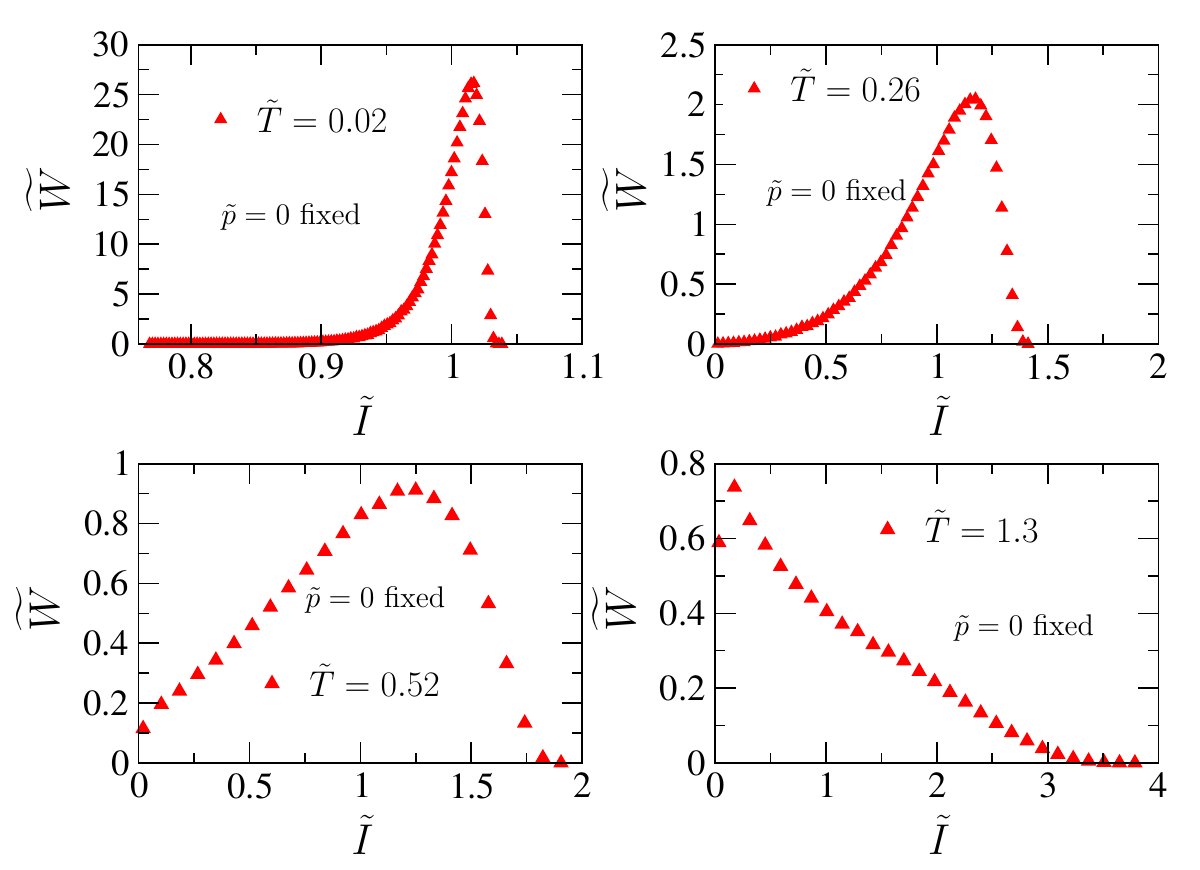}
\caption{Finite temperature distribution  of the normalized zero momentum intensity  $\widetilde{I}_{p=0}$, 
for different dimensionless temperatures $\widetilde{T}=k_B T/(K\Delta)$. 
The PDF crosses over from the zero temperature Gumbel distribution, 
Eq. \eqref{gumbel}, to an exponential distribution,  
as a signature of the thermal depletion of the quasi-condensate by the thermally populated  $p\neq 0$ modes. 
The experimentally accessible  temperature range, $T\sim 30\, {\rm nK}\,-\,120\,{\rm nK}$ \cite{experiment}, 
corresponds to $\widetilde{T}\sim\,0.12\,-\,0.46$. 
We used  $N=3500$, and $L= 39 \,\mu{\rm m}$ (density $\rho=90\,\mu {\rm m}^{-1}$),  and a chemical potential $\mu/h=1.6\,{\rm kHz}$, 
implying  $K\approx 77$, $c\approx 2,7 \,{\rm mm/s}$ and $\xi_h/L\approx 0.007$ for  $^{87}{\rm Rb}$ atoms.
 We  assumed  $\Delta \ptilde/(2\pi)=0.1$, corresponding to a time of flight $t=1\,$s, and  a real space resolution $\Delta R=12\,\mu {\rm m}$, but 
 shorter times of flight can also be applied using a focusing method, yielding similar images.
}\label{fig:FiniteTDist}
\end{figure}

This behavior and the crossover scale  in Eq.~\eqref{Tcrossover} are deeply related 
to the structure of    correlations in a  finite temperature  Luttinger liquid. At $T=0$ temperature, a bosonic Luttinger liquid exhibits  
power law correlations at distances larger than the healing length~\cite{Cazalilla,Imambekov2}.  At finite temperatures, 
however, these power law correlations turn into an exponential decay beyond the thermal wavelength~\cite{Cazalilla}, where
\be
\langle\hat{\psi}^\dagger(x)\hat{\psi}(0)\rangle\approx\rho\,\left(\frac{2\xi_h}{\lambda_T}\right)^{1/2K} e^{-|x|/\xi_T},\phantom{nn} \text{for }\, |x|>\lambda_T\,.
\label{eq:expcorrel}
\ee
Notice that the  \emph{thermal correlation  length} $\xi_T$ appearing here (often denoted by $\lambda_T$ in 
the literature) is proportional to but not identical with the \emph{thermal wavelength of the sound modes,   denoted here by 
$\lambda_T =\hbar c/(\pi k_B T)$}; 
being influenced by the stiffness of the condensate, $\xi_T$ is larger by a factor of $2K$~\cite{densityripples}, 
 $$
\xi_T = 2\,K\,\lambda_T\,,
$$
implying that  $\xi_T$ can be several orders of magnitude larger than $\lambda_T$ in a weakly interacting condensate.
%
Notice that the product $Kc\sim \rho /m$ is independent of the interaction strength by Galilean invariance~\cite{Kc_Galilei}.
Thus  the correlation length  $\xi_T\sim \hbar^2 \rho/(m k_B T) $ is independent of the strength of 
interaction. It is precisely this length scale that   appears in Eq~\eqref{Tcrossover}, which can be re-expressed as 
$\tilde T = L/(\xi_T\pi^2) $. The condition $\tilde T\lesssim 1$ thus corresponds to the inequality
$$
L\lesssim \xi_T \;\pi^2
$$
ensuring that the phase of the condensate remains close to  uniform for sizeable segments of  gas.
As shown in Appendix~\ref{app:expectation_value}, the number of particles in the $p=0$ mode is also determined by this ratio, 
$\langle \hat N_0\rangle\approx N\, 2\xi_T/L$. Thus $\tilde T\lesssim 1$  also implies that 
at least about  20 \% of the particles remain in the homogeneous condensate.
As stated earlier in this section, this condition is independent of the interaction strength. Indeed,
although  the discussion above focused  on the weakly interacting limit, $K\gg 1$,
we observe a similar crossover to an exponential function even for strong interactions, 
for which  $\lambda_T\sim \xi_T$ (see Appendix \ref{app:suppfig}).

%

The exponential distribution emerging for $\widetilde{T}\gtrsim 1$  can be understood as a consequence 
of the thermal depletion of the condensate by  low energy $p\neq 0$ modes. Considering the latter naively 
as particle reservoirs leads to 
$$
{\rm Prob}(\hat{N}_{p=0}=n)\propto  e^{-\beta\mu_{\rm eff} n},
$$
with some effective chemical potential $\mu_{\rm eff}$, set by the population of low energy modes.


\begin{figure}[t!]
\includegraphics[width=\columnwidth]{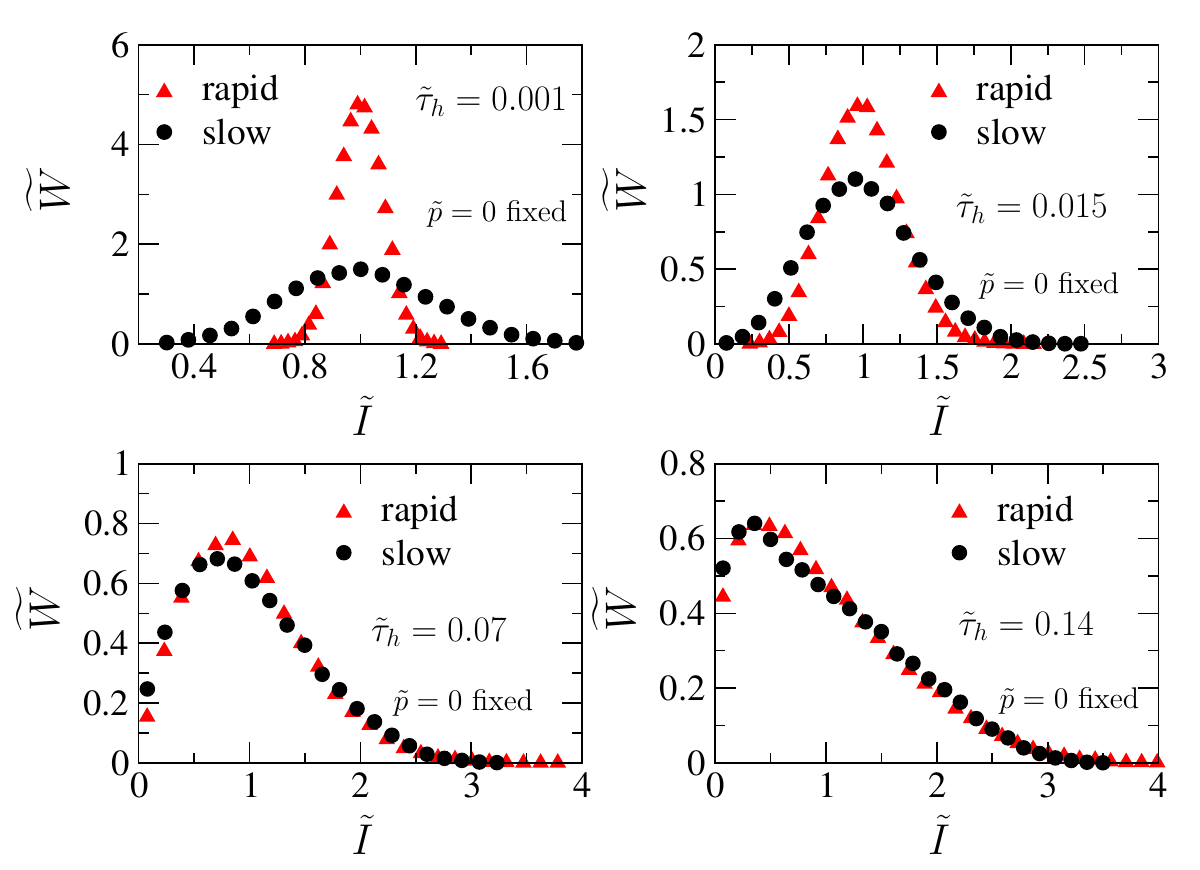}
\caption{ Distribution  of the normalized zero momentum intensity  $\widetilde{I}_{p=0}$ after an interaction quench, 
for different holding times, $\tau_h = 1.3\mu \rm{s} - 177 \mu \rm{s}$, measured in dimensionless units,    $\tilde{\tau}_h=\tau_h c_f/L$. Distributions are plotted for two different quenches of durations $\tau = 0.71\mu {\rm s}$ (rapid), and $\tau = 71\mu {\rm s}$ (slow). We have used $N=3684$, $L=39\mu \rm{ m}$ and a  chemical potential $\mu/h=1.6\,{\rm kHz}$, corresponding to  $K_0=80$, 
$c_0=2.7 \rm{mm/s}$, and  $\xi_h^0=0.27\mu \rm{m}$,    and assumed an interaction quench to  $K_f=7$, yielding $c_f=c_0 K_0/K_f = 30.8\; \rm{mm/s}$. 
We assumed a modest  momentum resolution, $\Delta\tilde{p}/(2\pi)=0.1$.
Similar to the finite temperature thermalization plotted in Fig. \ref{fig:FiniteTDist}, the PDF after a rapid quench crosses over from the equilibrium Gumbel distribution, 
Eq. \eqref{gumbel}, to an exponential distribution as $\tilde{\tau}_h$ increases, even though the number of excitations in the system remains 
constant after the quench. For a slower quench, the PDF for short holding times $\tau_h$ is much wider than the Gumbel distribution, showing that increasing interactions have time to deplete the quasi-condensate during the quench protocol, resulting in larger particle number fluctuations. As in the case of rapid quench, for larger holding times this PDF crosses over to a thermal distribution.  
}\label{fig:QuenchDist}
\end{figure}

\section{Distribution after interaction quenches}

So far we have focused on applying Time of Flight Full Counting Statistics to study equilibrium correlations.  
Even more interestingly, we can use it to study non-equilibrium dynamics and to
 gain information about the non-equilibrium states and time evolution of a system after a quantum quench \cite{ polkovnikovrmp,dziarmagareview}.

Here we demonstrate this perspective by  focusing on interaction quenches,  i.e., on changing  $g$  using 
a Feshbach resonance ~\cite{ZwergerReview}. For the sake of simplicity, we consider linear quench procedures of $g$, 
where the product $c(t)K(t) = \hbar \pi \rho/m$ remains constant by Galilean invariance~\cite{Kc_Galilei}, while 
$c /K$ changes approximately linearly  over a quench time $\tau$
\cite{footnote3}.
After the quench,  the atoms are held in the trap for an additional holding time $\tau_h$, while the final parameters $c_f$ and $K_f$ remain constant, and  the ToF experiment is performed only afterwards.

For short enough quench times $\tau$, the quench creates abundant excitations. Here we focus on these excitations and 
 concentrate therefore on zero temperature quenches. The initial state  is  then simply the Gaussian ground state  wave function 
 corresponding to the initial parameters $c_0$ and $K_0$.  Moreover, the wave function remains Gaussian 
 during the time evolution~\cite{quenchGritsev}, and can be expressed as 
\begin{equation*}
\Psi\left(\{\phi_k\},t\right)\sim \prod_{k>0} \exp\left(-\sigma_k(t)\,\phi_k\phi_{-k}\right),
\end{equation*}
with the  parameters $\sigma_k(t)$ obeying simple differential equations ~\cite{quenchGritsev}. This observation 
allows us to evaluate the full distribution of the intensity $\tilde{I}_p$, by only slightly modifying the derivation outlined in Appendix \ref{app:pdf}.

Fig. \ref{fig:QuenchDist} shows the intensity distribution of the zero mode, $\tilde{I}_0$, for a large quench between Luttinger parameters 
$K_0=80$ and $K_f=7$, as a function of the holding time after the quench, $\tau_h$. The distributions are plotted for two different quench times $\tau$. 

After a rapid quench, for short holding times the probability density function still resembles the Gumbel distribution, Eq.\eqref{gumbel}, valid for the 
equilibrium case. However,  we observe a crossover to an exponential distribution 
upon increasing the holding time, $\tau_h$. 
The phenomenon  observed  is similar to the finite temperature thermalization plotted in Fig. \ref{fig:FiniteTDist}, even though 
the number of excitations in each mode $k$ is a conserved quantity for the Luttinger model 
considered here, and the final state is definitely not thermal.

The structure of this non-thermal final state can be understood as follows. After long enough holding times, $\tau_h$, the distribution of the particle number $\hat{N}_p$ looks thermal for each momentum $p$. Based on this thermal, exponential distribution of $\hat{N}_p$, one can define an effective inverse temperature $\beta_p$ \cite{workstatistics},
\begin{equation*}
{\rm Prob}(\hat{N}_p=n)\propto e^{-\beta_p\varepsilon_p n},
\end{equation*}
with $\varepsilon_p$ denoting the quasiparticles'  dispersion relation. The non-thermal nature of the state  is reflected by the fact that 
in contrast to a thermal state characterized by a single inverse temperature,   $\beta_p$   strongly depend on the momentum $p$~\cite{workstatistics}. 
Similar pre-thermalization  phenomena are encountered in some quench experiments on closed, cold atomic systems, 
where the long-time expectation value of local observables can be well described by a thermal ensemble, despite the non-equilibrium state of the system \cite{greiner2,thermalization2}. 

For a slower quench, the distribution for short holding times $\tau_h$ gets much wider compared to the distribution after a sudden quench. 
This widening can be understood by noting that the interactions increase during the quench protocol. For slower quenches these stronger 
interactions have time to deplete the quasi condensate while the quench is performed, manifesting in more pronounced particle number 
fluctuations for short holding times. For larger holding times, however, we observe a crossover to a thermal distribution, similarly to the case of a rapid quench.

For both quench procedures, the time scale of thermalization of the  zero-mode is very fast, and for realistic parameters it falls to the range of  $\sim 0.1\;\rm{ms}$.
 


\section{Conclusions} 

 In this work we have proposed  a novel approach to analyse time of flight images, namely to measure the \emph{full probability distribution 
function} (PDF) of the intensities in a series of images. Similar to full counting statistics~\cite{levitov,nazarov}, 
the PDF of the intensity contains information on the complete distribution of the number of particles $N_p$ 
with a given momentum $p$, beyond its expectation value and variance, and reveals the structure of the quantum state
observed and its quantum fluctuations. This so far unexploited information in ToF images reflects the emergent universal behavior of strongly correlated low dimensional quantum systems.

We have demonstrated the perspectives of this versatile method on the specific example of an  interacting one-dimensional condensate.  
We have first focused on the equilibrium signal, and have shown that 
the intensity distribution of the image for $p\ne 0$ has an exponential character (deformed into a 
Gamma distribution with decreasing resolution), reflecting the squeezed structure  of the superfluid 
ground state. The $p=0$ intensity distribution, on the other hand, reveals fluctuations of the 
quasi-condensate, and turns out to be a Gumbel distribution in the weakly interacting limit, a familiar universal distribution from extreme value statistics. 
We have shown that the Gumbel distribution  derives from  particle number conservation, combined with 
large, interaction induced quantum fluctuations of the small momentum modes.

We have also shown that these intriguing  fingerprints of quantum fluctuations 
remain observable in a finite system at small but finite temperatures within the experimentally 
accessible range, but the predicted Gumbel   distribution is destroyed once the 
small momentum thermal modes thermalize the $p=0$ quasi-condensate mode.

ToF full counting statistics can be used in a versatile way to study non-equilibrium dynamics 
and thermalization.  As an example, we considered an interaction quench, and 
have shown that the intensity statistics of the $p=0$ mode displays clear signatures of 
(pre-)thermalization as a function of the holding time after the quench, whereby the 
original Gumbel distribution, discussed above turns into a quasi-thermal exponential (Gamma) 
distribution. This universal exponential distribution describes a condensate connected to a particle reservoir, formed by the $p>0$ modes.

One can also go beyond measuring the PDF of the intensity at a given point of the ToF image
by measuring the complete \emph{joint distribution functions}, $W(I_p,I_{p'})$,
rather than measuring just  intensity correlations, $\langle I_p I_{p'}\rangle$. 
As an example, we have determined  this joint distribution  for the $p=0$ quasi-condensate intensity and the $p\ne 0 $
intensities, and have shown that  $W(I_0,I_{p'})$   exhibits  strong negative correlations,
 induced by particle number conservation.  Clearly, our analysis can be generalized  to 
 multipoint distributions, $W(\{I_p\})$, still expected to reflect universality, though the experimental and theoretical 
 accessibility  becomes  less obvious  for these complex  quantities. 

As demonstrated here through the simplest example, ToF full counting statistics is expected to give insight to the exotic quantum states 
of various interacting quantum systems. Besides investigating the emergent universal behavior of low dimensional quantum systems, \emph{Time of Flight Full Counting Statistics} could also be applied to study  exotic quantum states  in higher dimensional, 
  fermionic or even anyonic  systems where it is supposed 
 to reflect the quantum statistics of particles. 
Another interesting direction would be the analysis of ToF full counting statistics at quantum critical points, such as the quantum 
 critical points of the transverse field Ising model or that of spinor condensates~\cite{spinorreview}, e.g., 
 where  quantum fluctuations get stronger and bare particles cease to exist. It is also a completely open question, how  ToF distributions reflect 
the structure of a  many-body localized state,  but the images of chaotic 
and integrable models are also expected to exhibit  different ToF
full counting statistics.  

\begin{acknowledgments}
This research has been  supported by the Hungarian Scientific  Research Funds Nos. K101244, K105149, SNN118028, K119442. 
ED acknowledges support from Harvard-MIT CUA, NSF Grant No. DMR-1308435, AFOSR Quantum Simulation MURI, AFOSR MURI Photonic Quantum Matter, the Humboldt Foundation, and the Max Planck Institute for Quantum Optics.
\end{acknowledgments}

\appendix

\section{Probability density function}
\label{app:pdf}

Here we derive the probability density function of the intensity, $W_p(I)$, both for the zero temperature case and for finite temperatures. First we perform the calculations at $T=0$, then we generalize the results to finite temperatures.

In order to derive the PDF at $T=0$, stated in Eqs. \eqref{w} and \eqref{gtf}, we have to calculate the momenta $\langle\hat{I}_{R,\Delta R}^n\rangle(t)$ for all $n$. First we express the intensity $\hat{I}_{R,\Delta R}(t)$ in terms of the  field operators at $t=0$ by substituting the free propagator $G(x,t)=\sqrt{\frac{m}{2\pi i t}}\,{\rm exp}(i m x^2/(2 t))$ into Eq. \eqref{def}, and use the density-phase representation \eqref{densphase} to arrive at
\begin{align}\label{t0}
 \hat{I}_{R,\Delta R}(t)=&\rho\dfrac{m\Delta R}{\sqrt{2 \pi}\,t}\int_{-L/2}^{L/2}{\rm d} x_1\int_{-L/2}^{L/2}{\rm d} x_2\,e^{-\frac{m^2\,\Delta R^2}{2\,t^2}(x_1-x_2)^2} \nonumber\\
& e^{\frac{i\,m R}{t}(x_1-x_2)-\frac{i m}{2\,t}(x_1^2-x_2^2)}e^{-i(\hat{\phi}(x_1,0)-\hat{\phi}(x_2,0))}.
\end{align}

The $n$th momentum of $\hat{I}_{R,\Delta R}(t)$ involves the $2n$ point correlator of the phase operator. This can be determined by using the Fourier expansion of $\hat{\phi}$, for open boundary conditions given by
\begin{align}\label{fourier}
&\hat{\phi}(x)=\frac{1}{\sqrt{L}}\hat{\phi}_0+\nonumber\\
&\quad\sum_{k>0}\sqrt{\dfrac{\pi}{K L|k|}}e^{-\xi_h|k|/2}\cos(k (x+L/2))\left(\hat{b}_k+\hat{b}_k^\dagger\right),
\end{align}
with $k=\pi j/ L $, $j\in\mathbb{Z}^+$. Here $\hat{b}_k^\dagger$ and $\hat{b}_k$ are bosonic creation and annihilation operators, with $\hat{b}_k$ annihilating the ground state of the system. The inverse of the healing length $\xi_h$ serves as a momentum cutoff. All ground state expectation values $\langle\hat{I}_{R,\Delta R}^n\rangle(t)$ can be calculated by using the normal ordering identity
\begin{equation*}
e^{D_k \hat{b}_k+D_k^*\hat{b}_k^\dagger}=e^{D_k^*\hat{b}_k^\dagger}e^{D_k \hat{b}_k}e^{-|D_k|^2/2}
\end{equation*}
with $D_k=\sqrt{\pi/(K L|k|)}e^{-\xi_h|k|/2}\cos(k (x+L/2))$, leading to
\begin{align}\label{moment}
&\langle  \hat{I}_{R,\Delta R}^n\rangle(t)=\left(\dfrac{L\rho\,\Delta\tilde{p}}{\sqrt{2\pi}}\right)^n\int...\int_{-1/2}^{1/2}\prod_{i=1}^{n}({\rm d} u_i{\rm d}v_i\,C(u_i,v_i))\cdot\nonumber\\
&\qquad\exp\left(-\sum_{j>0}\frac{e^{-\xi_h\pi j/L}}{2Kj}\left[\sum_{i=1}^n\left\lbrace\cos\left(\pi j u_i+\frac{j\pi}{2}\right)\right.\right.\right. \nonumber\\
&\qquad\qquad\qquad\quad-\left.\left.\left.\cos\left(\pi j v_i+\frac{j\pi}{2}\right)\right\rbrace\right]^2\right),
\end{align}
with 
\begin{equation}\label{Cuv}
C(u,v)=e^{ -\frac{\Delta\tilde{p}^2}{2}(u-v)^2+i\,\tilde{p}(u-v)\left(1-\frac{u+v}{2 R/L}\right)},
\end{equation}
and dimensionless variables $\tilde{p}$ and $\Delta\tilde{p}$ given by Eq. \eqref{pdp}.

The quadratic sum appearing in the exponent of Eq. \eqref{moment} can be decoupled by applying the Hubbard-Stratonovich transformation, performed by introducing a new integration variable $\tau_j$ for every index $j$,
\begin{align*}
&\exp\left(-\frac{e^{-\xi_h\pi j/L}}{2Kj}\left[\sum_{i=1}^n\left\lbrace\cos\left(\pi j u_i+\frac{j\pi}{2}\right)\right.\right.\right. \nonumber\\
&\qquad\qquad\qquad\quad-\left.\left.\left.\cos\left(\pi j v_i+\frac{j\pi}{2}\right)\right\rbrace\right]^2\right)=\\
&\int_{-\infty}^{\infty}\dfrac{{\rm d}\tau_j}{\sqrt{2\pi}}e^{-\tau_j^2/2}\exp\left(i\,\tau_j\frac{e^{-\xi_h\pi j/(2L)}}{\sqrt{Kj}}\times\right.\\
&\qquad\left.\sum_{i=1}^n\left\lbrace\cos\left(\pi j u_i+\frac{j\pi}{2}\right)-\cos\left(\pi j v_i+\frac{j\pi}{2}\right)\right\rbrace\right).
\end{align*}
By substituting this expression into Eq. \eqref{moment}, the integrals over different pairs of variables $\{u_i,v_i\}$ can be performed independently, and we arrive at
\begin{equation*}
\langle  \hat{I}_{R,\Delta R}^n\rangle(t)=\left(\dfrac{L\rho\,\Delta\tilde{p}}{\sqrt{2\pi}}\right)^n\int_{-\infty}^{\infty}\prod_{j>0}\dfrac{{\rm d}\tau_j}{\sqrt{2\pi}}\,e^{-\tau_j^2/2}g\left(\lbrace \tau_j\rbrace\right)^n,
\end{equation*}
with $g\left(\lbrace \tau_j\rbrace\right)$ given by Eq. \eqref{gtf}. Comparing this result with Eq. \eqref{pdfdef} shows, that the distribution of $\hat{I}_{R,\Delta R}(t)$ can indeed be described by a PDF, given by Eqs. \eqref{w} and \eqref{gtf}.

Now we generalize these results to $T>0$ temperatures. The Fourier expansion of the phase operator, Eq. \eqref{fourier}, together with the thermal occupation of the modes, $\langle\hat{b}_k^\dagger\hat{b}_k\rangle=1/(e^{\beta c k}-1)$, implies
\begin{align*}
&\langle e^{i\hat{\phi}(x)-i\hat{\phi}(y)}\rangle=\exp\left(-\sum_{j>0}\frac{e^{-\xi_h\pi j/L}}{2K\tanh(\beta c\pi j/(2L))}\times\right.\nonumber\\
&\qquad\quad\left.\left[\cos\left(\pi j x+\frac{j\pi}{2}\right)-\cos\left(\pi j y+\frac{j\pi}{2}\right)\right]^2\right).
\end{align*}
The only difference compared to the expectation value at $T=0$ temperature is the appearance of the thermal occupation factor $\tanh(\beta c\pi j/(2L))$. By repeating the derivation above, we find that the distribution function still takes the form Eq. \eqref{w}, but with a modified function $g_T\left(\lbrace \tau_j\rbrace\right)$ given by
\begin{align*}
&g_T\left(\lbrace \tau_j\rbrace\right)=\int\int_{-1/2}^{1/2}{\rm d}u\,{\rm d}v\,C(u,v)\times\\
&\qquad\exp\left(i\sum_{j}\tau_{j}\;\frac{e^{-\xi_h\pi {j}/(2L)}}{\sqrt{K\,j\tanh(\frac{\beta c\pi j}{2L})}}\left\lbrace\cos\left(\pi \,{j} \,u+\frac{{j}\,\pi}{2}\right)\right.\right. \nonumber\\
&\qquad\qquad\qquad\quad-\left.\left.\cos\left(\pi\, {j}\, v+\frac{{j}\,\pi}{2}\right)\right\rbrace\right).
\end{align*}

\section{Joint distribution function}
\label{app:jointpdf}

In this appendix we derive a numerically tractable expression for the joint PDF at $T=0$ temperature, defined in Eq. \eqref{jointpdf}, by calculating the momenta $\langle\hat{I}_{1}^{n_1}\hat{I}_{2}^{n_2}\rangle(t)$ for all $n_1$ and $n_2$. Here we introduced the shorthand notation $\hat{I}_{1}\equiv\hat{I}_{R_1,\Delta R_1}$. By using Eq. \eqref{t0} and the Fourier expansion of the phase operator, Eq. \eqref{fourier}, we arrive at
\begin{align}\label{jointmoment}
&\langle  \hat{I}_{1}^{n_1}\hat{I}_{2}^{n_2}\rangle(t)=\left(\dfrac{L\rho}{\sqrt{2\pi}}\right)^{n_1+n_2}\Delta\tilde{p}_1^{n_1}\Delta\tilde{p}_2^{n_2}\nonumber\\
&\int...\int_{-1/2}^{1/2}\prod_{i=1}^{n_1}({\rm d} u_{i}{\rm d}v_{i}\,C_1(u_{i},v_{i}))\prod_{l=1}^{n_2}({\rm d} \tilde{u}_{l}{\rm d}\tilde{v}_{l}\,C_2(\tilde{u}_{l},\tilde{v}_{l}))\cdot\nonumber\\
&\exp\left(-\sum_{j>0}\frac{e^{-\xi_h\pi j/L}}{2Kj}\left[\sum_{i=1}^{n_1}\left\lbrace\cos\left(\pi j u_i+\frac{j\pi}{2}\right)\right.\right.\right. \nonumber\\
&-\left.\left.\left.\cos\left(\pi j v_i+\frac{j\pi}{2}\right)\right\rbrace+\sum_{l=1}^{n_2}\left\lbrace\cos\left(\pi j \tilde{u}_l+\frac{j\pi}{2}\right)\right.\right.\right.\nonumber\\
&\qquad\qquad\left.\left.\left.-\cos\left(\pi j \tilde{v}_l+\frac{j\pi}{2}\right)\right\rbrace\right]^2\right),
\end{align}
with $C_i(u,v)$ given by Eq. \eqref{Cuv} with parameters $R_i$ and $\Delta R_i$ for $i=1,2$.

Similarly to the calculation presented in Appendix \ref{app:pdf}, the quadratic sum appearing in the exponent of Eq. \eqref{jointmoment} can be decoupled by applying a Hubbard-Stratonovich transformation. By introducing a new integration variable $\tau_j$ for every index $j$, and performing the integrals over different pairs of variables $\{u_i,v_i\}$ and $\{\tilde{u}_l,\tilde{v}_l\}$ independently, we arrive at
\begin{align*}
\langle  \hat{I}_{1}^{n_1}\hat{I}_{2}^{n_2}\rangle(t)&=\left(\dfrac{L\rho}{\sqrt{2\pi}}\right)^{n_1+n_2}\Delta\tilde{p}_1^{n_1}\Delta\tilde{p}_2^{n_2}×\nonumber\\
&\int_{-\infty}^{\infty}\prod_{j>0}\dfrac{{\rm d}\tau_j}{\sqrt{2\pi}}\,e^{-\tau_j^2/2}g_1\left(\lbrace \tau_j\rbrace\right)^{n_1}g_2\left(\lbrace \tau_j\rbrace\right)^{n_2},
\end{align*}
with $g_i\left(\lbrace \tau_j\rbrace\right)$ given by Eq. \eqref{gtf} with parameters $R_i$ and $\Delta R_i$ for $i=1,2$. Comparing this expression to the definition of the joint distribution, Eq. \eqref{jointpdf}, we find that
\begin{align*}
&W(I_1,I_2)=\int\int_{-\infty}^{\infty}\prod_{j}\dfrac{{\rm d}\tau_{j}\,e^{-\tau_{j}^2/2}}{\sqrt{2\pi}}\times\\
&\delta\left(I_1-\dfrac{N\Delta \widetilde{p_1}}{\sqrt{2\pi}}g_1\left(\lbrace \tau_{j}\rbrace\right)\right)\delta\left(I_2-\dfrac{N\Delta \widetilde{p_2}}{\sqrt{2\pi}}g_2\left(\lbrace \tau_{j}\rbrace\right)\right).
\end{align*}
The distribution can be evaluated by performing a Monte Carlo simulation for the normal random variables $\tau_j$, and calculating the two dimensional histogram for $I_1$ and $I_2$.

\section{Focusing technique}
\label{app:focusing}

Besides the ToF measurements, the focusing technique provides an alternative way to access the momentum distribution~\cite{focusing0,focusing1,focusing}. 
The strong transverse confinement of the quasi one dimensional system is abruptly switched off, while the weak longitudinal confinement is replaced by a strong harmonic trap of frequency $\omega$~\cite{footnote3}, and the gas is imaged after a quarter time period, $t=T/4=\pi/(2\omega)$.

To express the intensity \eqref{def} in this case with the field operators at time $t=0$, we have to replace the free 
propagator in Eq. \eqref{eq:propagator} by that  of the harmonic oscillator
$G_{\rm osc}(x,y,t=T/4)={e^{-i\,x\, y/l_0^2}}/({l_0\sqrt{2\pi\hbar\, i}})$, 
with $l_0=\sqrt{\hbar/(m\omega)}$  the oscillator length of the strong trapping potential. 
In this case,  Eqs.~\eqref{eq:propagator}  thus simply yields   the Fourier transform of the field at $t=0$, 
$$
\hat{\psi}(R,t=T/4)\sim \hat{\psi}_p
$$
at a momentum $p= \hbar R/l_0^2$. Thus the intensity measured at $R$ is directly proportional to the number of particles $\hat{N}_p$ in this case. 
 Performing  calculations similar to those sketched in Appendix \ref{app:pdf}, 
we arrive at  Eqs. ~\eqref{w} and Eq.~\eqref{gtf}, with the weight function ~\eqref{Cuv} replaced by
\begin{align*}
C(u,v)=\exp\left(-\dfrac{\Delta\tilde{p}_{\rm osc}^2}{2}(u-v)^2+i\,\tilde{p}_{\rm osc}(u-v)\right).
\end{align*}
and the 
dimensionless momentum and momentum resolution expressed as 
$\tilde{p}_{\rm osc}= {R\, L}/{l_0^2}$ and $ \Delta\tilde{p}_{\rm osc}= {\Delta R\;L}/{l_0^2}$.
Apart from  these  minor corrections, all our  calculations can be performed for focusing experiments, 
and while this method allows to use shorter measurement times,  the conclusions in the main text remain unaltered. 

\section{The expectation value of $\hat{I}_{R,\Delta R}(t)$}
\label{app:expectation_value}

In this appendix we investigate the expectation value of the intensity $\hat{I}_{R,\Delta R}(t)$, scaled out from the distribution functions calculated in the main text.

\begin{figure}[t!]
\includegraphics[width=\columnwidth]{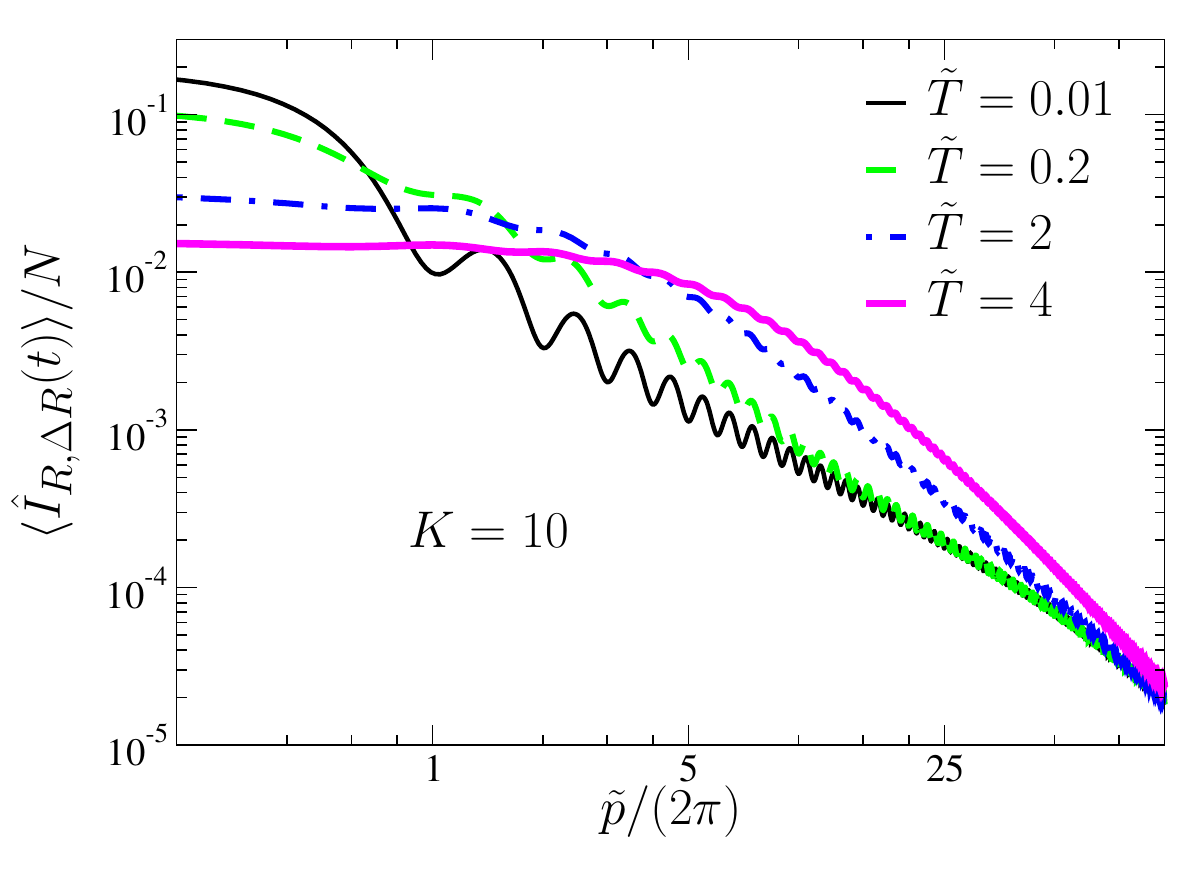}
\caption{Expectation value $\langle\hat{I}_{R,\Delta R}(t)\rangle/N$ plotted as a function of dimensionless wave number $\tilde{p}/(2\pi)$, for different dimensionless temperatures $\tilde{T}$, with parameters $K=10$, $\xi_h/L=0.002$ and $\Delta \tilde{p}=0.1\times 2\pi$, using logarithmic scale on both axis. As $T$ increases, the expectation value develops a wide flat region for small momenta $p\ll \hbar/\xi_T$. For larger momenta $\hbar/\xi_T\ll p\ll \hbar/\lambda_T$, the intensity shows a power law decay $\sim 1/p^2$. This behaviour can be explained by the exponential decay of two-point correlations in finite temperature Luttinger-liquids, Eq. \eqref{eq:expcorrel}, with correlation length $\xi_T$. For even larger momenta $p\gg \hbar/\lambda_T$, we get back the zero temperature results, resulting in a crossover to the different power law behavior  $\langle\hat{I}_{R,\Delta R}(t)\rangle\sim 1/p$.}\label{fig:expFiniteT}
\end{figure}

In order to investigate the temperature dependence of the expectation value, we plotted $\langle\hat{I}_{R,\Delta R}(t)\rangle/N$ as a function of dimensionless momentum $\tilde{p}/(2\pi)$ for different dimensionless temperatures $\tilde{T}$ in Fig. \ref{fig:expFiniteT}. We concentrated on the weakly interacting regime, keeping the Luttinger-parameter, $K=10$, constant. The low temperature results show pronounced oscillations, originating from the presence of the quasi-condensate due to finite size effects. For higher temperatures, the intensity $\langle\hat{I}_{R,\Delta R}(t)\rangle$ increases for non-zero momenta $\tilde{p}=O(2\pi)$, while the zero-momentum expectation value decreases due to the depletion of the condensate. Moreover, we can distinguish two momentum regions, corresponding to different behavior of the expectation value. For momenta much smaller than the thermal wavelength, $p\ll \hbar/\lambda_T$ (or $\tilde{p}/(2\pi)\ll \pi K\tilde{T}$ in dimensionless variables), the expectation value of the intensity is well approximated by the Fourier transform of Eq. \eqref{eq:expcorrel}, yielding
\begin{equation*}
\langle\hat{N}_p\rangle\approx N\,\left(\frac{2\xi_h}{\lambda_T}\right)^{1/2K}\dfrac{2\xi_T/L}{1+(p\,\xi_T/\hbar)^2}.
\end{equation*}
This expression predicts a power law decay $\langle\hat{I}_{R,\Delta R}(t)\rangle\sim 1/p^2$ for momenta $\hbar/\xi_T\ll p\ll \hbar/\lambda_T$. However, for even larger momenta, $p\gg \hbar/\lambda_T$, the short distance behaviour of the correlation function $\langle\hat{\psi}^\dagger(x)\hat{\psi}(0)\rangle$ becomes important, and it is not appropriate to approximate it by the simple exponential function Eq. \eqref{eq:expcorrel}. In this region the expectation value of the intensity converges to the zero temperature result, corresponding to a different power law behavior $\langle\hat{I}_{R,\Delta R}(t)\rangle\sim 1/p^{1-1/2K}\approx 1/p$.

This crossover between different power law decays is only observable in the limit of weak interactions, where $\lambda_T\ll \xi_T$, thus $\hbar/\xi_T\ll p\ll \hbar/\lambda_T$ is satisfied in a wide momentum range. In this case the Bogoliubov approximation is also valid, thus the same $\sim 1/p^2$ decay can also be explained by applying the Bogoliubov approach.

As expected, Bogoliubov theory is also able to account for the cross-over discussed above. According to Eq. \eqref{Bogoliubov}, the zero temperature Bogoliubov calculation gives $\langle\hat{N}_p\rangle\sim 1/|p|$. This result can be generalized to finite temperatures by including the appropriate Bose function, and taking into account the low energy dispersion relation $\varepsilon_p=c|p|$, resulting in
$$\langle\hat{N}_p\rangle\sim\dfrac{\coth(\beta c|p|/2)}{|p|}\sim 1/p^2.$$
Here the last approximation is valid for $p\ll 2 k_B T/ c\approx \hbar/\lambda_T$. As already mentioned, this $\sim 1/p^2$ decay is consistent with the numerical results plotted in Fig. \ref{fig:expFiniteT}.

\section{Gumbel distribution}
\label{app:gumbel}

In this appendix we show that the Gumbel distribution \eqref{gumbel}, arising for weak interactions, can be derived from the structure of the Bogoliubov ground state, by taking into account particle number conservation. In this perturbative approach, the PDF \eqref{gumbel} emerges as the distribution of the normalized operator giving the number of particles with zero momentum, 
$$\tilde{N}_0=\dfrac{\hat{N}_{p=0}-\langle\hat{N}_{p=0}\rangle}{\delta N_{p=0}}.$$
Here $\langle\hat{N}_{p=0}\rangle$ denotes the expectation value, and $\delta N_{p=0}$ is the standard deviation. For simplicity, we perform the calculations using periodic boundary conditions.

As already noted in the main text, particle number conservation implies
\begin{equation}\label{Nconserv}
\hat{N}_{p=0}=N-\sum_{p\neq 0}\hat{N}_p,
\end{equation}
with $N$ denoting the total number of particles. Moreover, the two mode squeezed structure of the ground state in non-zero momenta $p$ and $-p$, resulting in a perfect correlation $\hat{N}_p=\hat{N}_{-p}$, leads to an exponential distribution for the random variable $(\hat{N}_p+\hat{N}_{-p})/N$, with expectation value $\hbar\pi/(K L|p|)$ \cite{Bogoliubov} (see Eq. \eqref{Bogoliubov}). For PBC the momentum can only take values $p=2\pi n \hbar/L$, so the PDF of the sum $\sum_{p\neq 0}\hat{N}_p/N$ can be written as
\begin{align}\label{sumPDF}
&P\left(\sum_{p\neq 0}\hat{N}_p/N=x\right)=\prod_{i=1}^{n_c}(2Kn)\int_0^\infty{\rm d}x_1\,e^{-2K\,x_1}\times\nonumber\\
&\int_0^\infty{\rm d}x_2\,e^{-2K\, 2\,x_2}...\int_0^\infty{\rm d}x_{n_c}e^{-2K\,n_c\, x_{n_c}}\delta\left(x-\sum_{i=1}^{n_c}x_i\right).
\end{align}
Here $n_c\sim L/\xi_h$ denotes a cutoff in momentum space, restricting the momentum $p$ to the low energy region, described by linear dispersion relation.

The PDF \eqref{sumPDF} can be rewritten by introducing new integration variables $z_1=x_{n_c}$, $z_2=x_{n_c}+x_{n_c-1}$, ..., and $z_{n_c}=\sum_{i=1}^{n_c}x_i$ as
\begin{align*}
&P\left(\sum_{p\neq 0}\hat{N}_p/N=x\right)=(2K)^{n_c}\,n_c!\,\times\\
&\int_0^\infty{\rm d}z_1\int_{z_1}^\infty{\rm d}z_2...\int_{z_{n_c-1}}^\infty{\rm d}z_{n_c}e^{-2K\sum_{i=1}^{n_c}z_i}\,\delta\left(x-z_{n_c}\right).
\end{align*}
This result shows, that the PDF associated to the operator $\sum_{p\neq 0}\hat{N}_p/N$ is equivalent to the distribution of the maximum of $n_c$ independent, exponentially distributed random variables, with equal expectation values $1/(2K)$. This observation follows from noting, that the integrand describes independent exponential random variables, subject to the constraint $z_1<z_2<...<z_{n_c}$, with the factor $n_c!$ taking into account all possible orderings of these $n_c$ variables. This interpretation explains the emergence of the extreme value distribution $W_{\rm Gumbel}$.

The cumulative distribution function of the maximum of independent random variables can be easily calculated, leading to the probability
\begin{align}\label{CDF}
{\rm Prob}\left(\tilde{N}_0<x\right)={\rm Prob}\left(\sum_{p\neq 0}\dfrac{\hat{N}_p}{N}>\sum_{p\neq 0}\dfrac{\langle\hat{N}_p\rangle}{N}-x\dfrac{\delta N_{p=0}}{N}\right)\nonumber\\
=1-\left(1-{\rm exp}\left\lbrace -2K\left(\sum_{p\neq 0}\dfrac{\langle\hat{N}_p\rangle}{N}-x\dfrac{\delta N_{p=0}}{N}\right)\right\rbrace\right)^{n_c}\nonumber\\
\approx 1-{\rm exp}\left(-n_c\,{\rm exp}\left\lbrace-2K\left(\sum_{p\neq 0}\dfrac{\langle\hat{N}_p\rangle}{N}-x\dfrac{\delta N_{p=0}}{N}\right)\right\rbrace\right),
\end{align}
with the approximation in the third line valid for large $K$. Here the expectation value $\sum_{p\neq 0}\langle\hat{N}_p\rangle/N$ is given by
\begin{equation*}
\dfrac{\langle\hat{N}_p\rangle}{N}=\sum_{n=1}^{n_c}\dfrac{1}{2Kn}.
\end{equation*}

Moreover, using the particle number conservation \eqref{Nconserv}, and the variances the variables $\hat{N}_{p\neq 0}$,  the standard deviation $\delta N_{p=0}/N$ can be calculated as
$$\dfrac{\delta N_{p=0}}{N}=\sqrt{\sum_{n=1}^{n_c}\left(\dfrac{1}{2Kn}\right)^2}\approx \dfrac{\pi}{2\sqrt{6}K},$$
taking the limit of large cutoff $n_c$ in the last step. Substituting these results into \eqref{CDF} allows us to take the $n_c\rightarrow\infty$ limit, resulting in the cumulative distribution function
\begin{equation*}
{\rm Prob}\left(\tilde{N}_0<x\right)\approx 1-{\rm exp}\left\lbrace-\exp\left(\dfrac{\pi}{\sqrt{6}}x-\gamma\right)\right\rbrace,
\end{equation*}
with $\gamma$ denoting the Euler constant, defined by the relation
\begin{equation*}
\gamma=\lim_{n_c\rightarrow\infty}\,\sum_{n=1}^{n_c}\dfrac{1}{n}-{\rm log}\, n_c.
\end{equation*}
By taking the derivative of this cumulative distribution function, we arrive at the PDF of the Gumbel distribution, Eq. \eqref{gumbel}.

\section{Numerical results for strong interactions}
\label{app:suppfig}

\begin{figure}[t!]
\includegraphics[width=0.9\columnwidth]{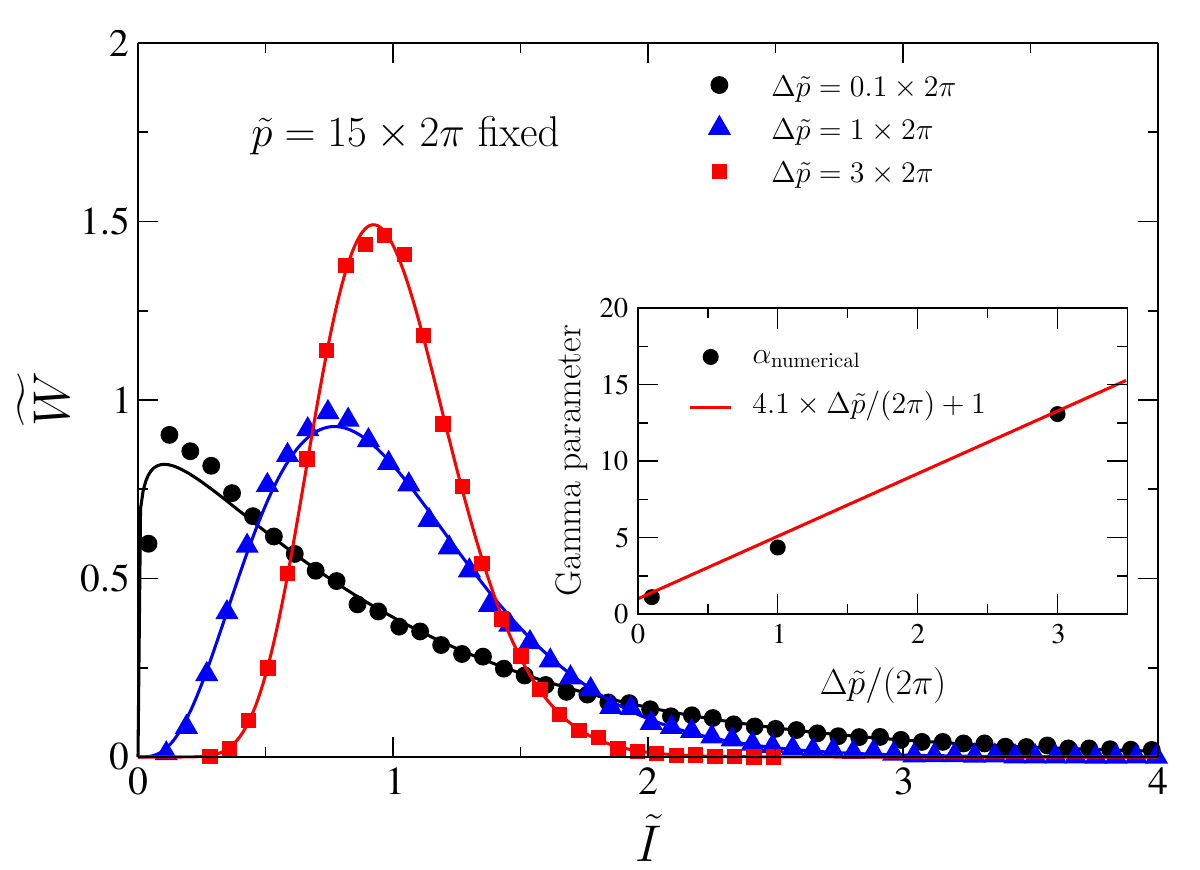}
\caption{Distribution  of normalized intensity $\widetilde{I}$ (symbols) at $T=0$ for stronger interactions, and fits with the Gamma distribution from Eq.~\eqref{Gamma} (solid lines), plotted for different momentum 
resolutions  $\Delta \widetilde{p}$. We used  $K=2$, $\widetilde{p}=15\times 2\pi$ and $\xi_h/L=0.002$.  Similarly to the limit of weak interactions, the distribution smoothly evolves from exponential to Gamma as $\Delta p$ increases. 
Inset: parameter of the fitted Gamma distribution $\alpha$ as a function of $\Delta \ptilde$, increasing approximately linearly with the same slope as in the weakly interacting limit.} 
\label{fig:gammasupp}
\end{figure}

\begin{figure}[t!]
\includegraphics[width=\columnwidth]{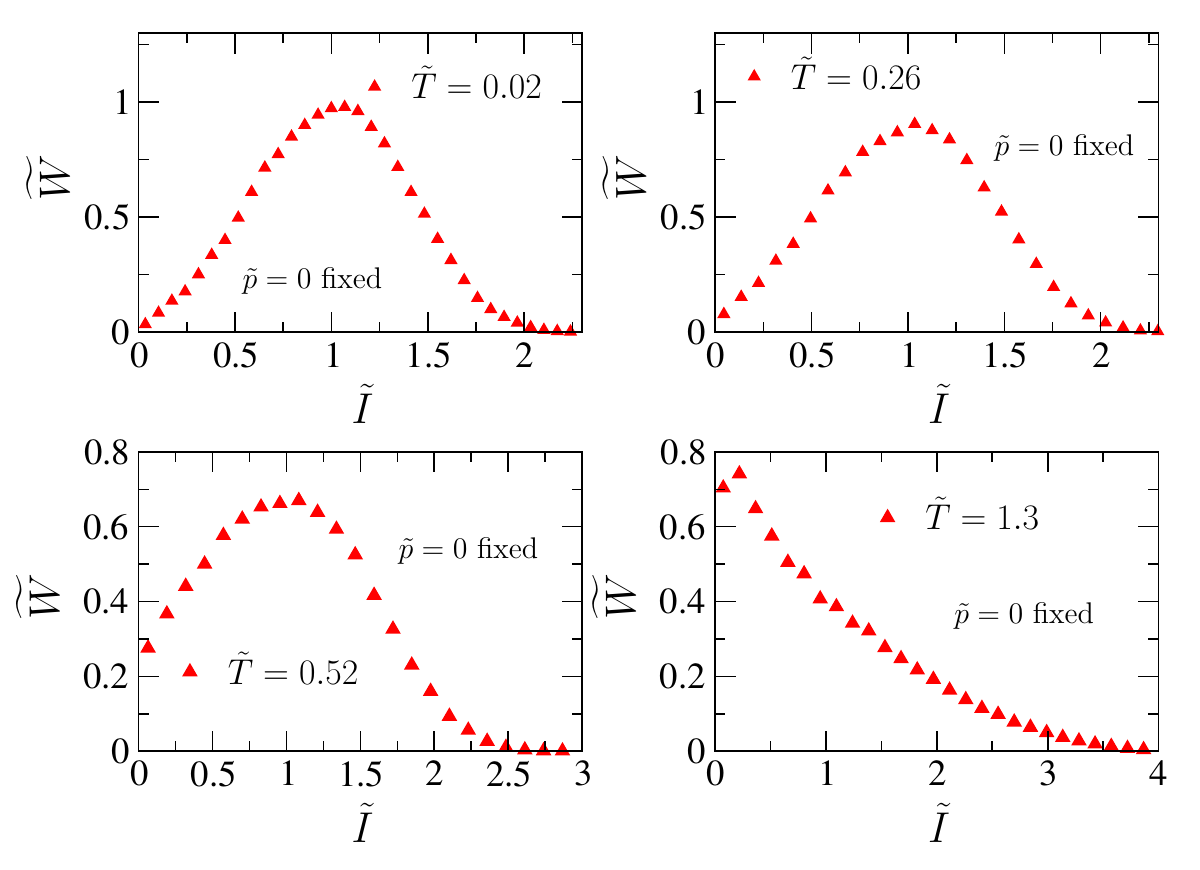}
\caption{Finite temperature distribution  of the normalized zero momentum intensity  $\widetilde{I}_{p=0}$ for strong interactions, 
using different dimensionless temperatures $\widetilde{T}=k_B T/(K\Delta)$. 
The PDF crosses over from the zero temperature limit (deviating from Gumbel distribution due to strong interactions) to an exponential distribution,  
as a signature of the depletion of the zero mode by the thermally populated  $p\neq 0$ modes. As in the limit of weak interactions, the crossover is governed by the dimensionless temperature $\tilde{T}$. We used  $K=1.5$,  $\Delta \ptilde/(2\pi)=0.1$ and $\xi_h/L\approx 0.002$.}
\label{fig:FiniteTDistsupp}
\end{figure}

In the figures of the main text we concentrated mostly on the limit of weak interactions. Here we present additional numerical results, corresponding to stronger interactions.

By analyzing the equilibrium quantum fluctuations at $T=0$ temperature, we have shown in Sec. \ref{sub:T0} that the distribution of the intensity at finite momentum crosses over from exponential to Gamma distribution with increasing momentum resolution $\Delta p$. We plotted this crossover for weak interactions in Fig. \ref{fig:gamma}. In Fig. \ref{fig:gammasupp} we show the same crossover for stronger interactions $K=2$. We find that the parameter of the fitted Gamma distribution, Eq.~\eqref{Gamma}, increases approximately linearly with $\Delta p$, with the same slope as in the limit of weak interactions. 

We considered the finite temperature distribution of the zero mode in Sec. \ref{sub:finiteT}. In the limit of weak interactions, plotted in Fig. \ref{fig:FiniteTDist} of the main text, we found a crossover from the zero temperature Gumbel distribution to an exponential distribution, as the temperature is increased and thermal fluctuations deplete the quasi-condensate. We observe a similar crossover for strong interactions $K=1.5$, by plotting the zero-momentum distributions for different dimensionless temperatures $\tilde{T}$ in Fig. \ref{fig:FiniteTDistsupp}. For such strong interactions, the distribution at $T=0$ deviates from the Gumbel distribution considerably (see also Fig. \ref{fig:k0} in the main text), but a clear crossover from the $T=0$ limit to an exponential distribution, governed by the dimensionless temperature $\tilde{T}$, still persists.


\begin{references}
\bibitem{footnote1}In case of simply taking a picture of the condensate, $\Delta R$
 would represent the  resolution of the optical system.
\bibitem{footnote2}Luttinger-liquid theory, being an effective long wave length description, loses its validity below the length scale $\xi_h$ \cite{Giamarchi}.
\bibitem{footnote3}In the perturbative limit, Eq.~\eqref{perturb}, the ratio  $c /K$    is directly proportional to the interaction $g$.
\bibitem{footnote4}The measurement of the momentum distribution in a ToF experiment requires long expansion times. The drawback of this method is that the much faster expansion of the transverse directions can result in a loss of signal. This problem can be circumvented by the focusing technique, allowing much shorter expansion times and better signal to noise ratios.
\bibitem{footnote5}In real experiments the atoms are confined to an optical trap, before being released to propagate freely during the ToF measurement. In our calculations we neglect the confining potential, and consider a homogeneous density with open or periodic boundary conditions for the phase operator. This approximation is justified, if the measured intensity is determined by the quantum fluctuations at the center of trap, where the density varies slowly, allowing a local density approximation.
\bibitem{footnote6}Noise correlations, extracted from time of flight images, have already been used as a versatile tool to study strongly correlated many-body systems, see Refs. \cite{noise,folling,fermionnoise}.
\end{references}
\end{document}